\documentclass{gtmon_a}
\pdfoutput=1

\usepackage{pinlabel}

\proceedingstitle{Groups, homotopy and configuration spaces (Tokyo
  2005)}
\conferencestart{5 July 2005}
\conferenceend{11 July 2005}
\conferencename{Groups, homotopy and configuration spaces,
                in honour of Fred Cohen's 60th birthday}
\conferencelocation{University of Tokyo, Japan}

\editor{Norio Iwase}
\givenname{Norio}
\surname{Iwase}

\editor{Toshitake Kohno}
\givenname{Toshitake}
\surname{Kohno}

\editor{Ran Levi}
\givenname{Ran}
\surname{Levi}

\editor{Dai Tamaki}
\givenname{Dai}
\surname{Tamaki}

\editor{Jie Wu}
\givenname{Jie}
\surname{Wu}

\title{Homotopy algebra of open--closed strings}                    

\author{Hiroshige Kajiura}
\givenname{Hiroshige}
\surname{Kajiura}
\address{Research Institute for Mathematical Sciences\\Kyoto University\\\newline 
Kyoto 606-8502\\Japan}
\email{kajiura@kurims.kyoto-u.ac.jp}
\urladdr{}

\author{Jim Stasheff}
\givenname{Jim}
\surname{Stasheff}
\address{Department of Mathematics\\University of Pennsylvania\\\newline
Philadelphia, PA 19104-6395\\USA}
\email{jds@math.upenn.edu}
\urladdr{}

\dedicatory{Dedicated to Fred Cohen in honor of his 60th birthday}

\volumenumber{13}
\issuenumber{}
\publicationyear{2008}
\papernumber{10}
\startpage{229}
\endpage{259}

\doi{}
\MR{}
\Zbl{}

\arxivreference{hep-th/0606283}

\keyword{homotopy algebra}
\keyword{moduli spaces of Riemann surfaces}
\keyword{deformation theory}
\subject{primary}{msc2000}{18G55}
\subject{secondary}{msc2000}{81T18}

\received{31 May 2006}
\revised{22 January 2007}
\accepted{25 January 2007}
\published{22 February 2008}
\publishedonline{22 February 2008}
\proposed{}
\seconded{}
\corresponding{}
\version{}

\makeatletter
\def\cnewtheorem#1[#2]#3{\newtheorem{#1}{#3}[section]
\expandafter\let\csname c@#1\endcsname\c@thm}

\let\xysavmatrix\xymatrix
\def\xymatrix{\disablesubscriptcorrection\xysavmatrix}
\AtBeginDocument{\let\bar\wbar\let\tilde\wtilde\let\hat\what}

\newtheorem{thm}{Theorem}[section]    
\cnewtheorem{lem}[thm]{Lemma}          
\cnewtheorem{prop}[thm]{Proposition}
\cnewtheorem{cor}[thm]{Corollary}

\theoremstyle{definition}
\cnewtheorem{defn}[thm]{Definition}    
\cnewtheorem{rem}[thm]{Remark}

\makeatother  

\makeautorefname{subsection}{subsection}
\makeautorefname{defn}{Definition}

\newcommand{\bp}{\begin{pmatrix}}
\newcommand{\ep}{\end{pmatrix}}

\newcommand{\bps}{\begin{smallmatrix}}
\newcommand{\eps}{\end{smallmatrix}}

\newcommand{\ti}{\tilde}

\def \cA{{\cal A}}

\def \cH{{\cal H}}
\def \cL{{\cal L}}
\def \cM{{\cal M}}
\def \cN{{\cal N}}
\def \cMC{{\cal MC}}
\def \cMb{{\bar {\cal M}}}
\def \cMub{\mbox{\underbar{${\cal M}$}}}

\def \cOC{{\cal OC}}

\def \f{{\frak f}}
\def \g{{\frak g}}
\def \l{{\frak l}}
\def \m{{\frak m}}
\def \n{{\frak n}}

\def \S{{\frak S}}

\def \cb{{\bar c}}

\def \lraw{\leftrightarrow}
\def \raw{\rightarrow}

\def \lgraw{\longrightarrow}

\def \End{\mathrm{End}}
\def \Hom{\mathrm{Hom}}
\def \Der{\mathrm{Der}}
\def \Coder{\mathrm{Coder}}

\def \deg{\mathrm{deg}}

\def \codim{\mathrm{codim}}

\def\ott{\otimes}

\def \ov#1{\frac{1}{#1}}

\def\b1{{\mathbb 1}}

\def \flpartial#1{\frac{\overleftarrow{\partial}}{\partial #1}}

\def \0{{\bf 0}}
\def \1{{\bf 1}}

\def \eb{{\bf e}}

\def \tri{\triangle}

\def \-{-\hspace*{-0.2cm}-}
\def \ie{ie}

\begin{document}

\begin{htmlabstract}
This paper is a survey of our previous works on open&ndash;closed homotopy
algebras, together with geometrical background, especially in terms of
compactifications of configuration spaces (one of Fred's specialities)
of Riemann surfaces, structures on loop spaces, etc.  We newly present
Merkulov's geometric A<sub>&infin;</sub>&ndash;structure [Internat. Math. Res.
Notices (1999) 153&ndash;164] as a special example of an OCHA.  We also
recall the relation of open&ndash;closed homotopy algebras to various
aspects of deformation theory.
\end{htmlabstract}

\begin{abstract}   
This paper is a survey of our previous works on open--closed homotopy
algebras, together with geometrical background, especially in terms of
compactifications of configuration spaces (one of Fred's specialities)
of Riemann surfaces, structures on loop spaces, etc.  We newly present
Merkulov's geometric $A_\infty$--structure \cite{mer2} as a special
example of an OCHA.  We also recall the relation of open--closed
homotopy algebras to various aspects of deformation theory.
\end{abstract}

\begin{webabstract}
This paper is a survey of our previous works on open--closed homotopy
algebras, together with geometrical background, especially in terms of
compactifications of configuration spaces (one of Fred's specialities)
of Riemann surfaces, structures on loop spaces, etc.  We newly present
Merkulov's geometric $A_\infty$--structure [Internat. Math. Res.
Notices (1999) 153--164] as a special example of an OCHA.  We also
recall the relation of open--closed homotopy algebras to various
aspects of deformation theory.
\end{webabstract}

\begin{asciiabstract}
This paper is a survey of our previous works on open-closed homotopy
algebras, together with geometrical background, especially in terms of
compactifications of configuration spaces (one of Fred's specialities)
of Riemann surfaces, structures on loop spaces, etc.  We newly present
Merkulov's geometric A_infty-structure [Internat. Math. Res.
Notices (1999) 153--164] as a special example of an OCHA.  We also
recall the relation of open-closed homotopy algebras to various
aspects of deformation theory.
\end{asciiabstract}

\maketitle

\section{Introduction}
\label{sec:1}

Open--closed homotopy algebras (OCHAs) (Kajiura and Stasheff
\cite{ocha}) are inspired by Zwiebach's open--closed string field
theory \cite{z:oc}, which is presented in terms of decompositions of
moduli spaces of the corresponding Riemann surfaces.  The Riemann
surfaces are (respectively) spheres with (closed string) punctures and
disks with (open string) punctures on the boundaries.  That is, from
the viewpoint of conformal field theory, classical closed string field
theory is related to the conformal plane $\C$ with punctures and
classical open string field theory is related to the upper half plane
$H$ with punctures on the boundary.  Thus classical closed string
field theory has an $L_\infty$--structure (Zwiebach \cite{z:csft},
Stasheff \cite{Sta1993}, Kimura, Stasheff and Voronov \cite{ksv1}) and
classical open string field theory has an $A_\infty$--structure
(Gaberdiel and Zwiebach \cite{GZ}, Zwiebach \cite{z:oc}, Nakatsu
\cite{nakatsu} and Kajiura \cite{Ka}).  The algebraic structure, we
call it an OCHA, that the classical open--closed string field theory
has is similarly interesting since it is related to the upper half
plane $H$ with punctures both in the bulk and on the boundary.

In operad theory (see Markl, Shnider, and Stasheff \cite{MSS:book}),
the relevance of the little disk operad to closed string theory is
known, where a (little) disk is related to a closed string puncture on
a sphere in the Riemann surface picture above.
The homology of the little disk operad defines a Gerstenhaber algebra
(Cohen \cite{cohen}, Getzler and Jones \cite{getzler-jones}), in
particular, a suitably compatible graded commutative algebra structure
and graded Lie algebra structure.  The framed little disk operad is in
addition equipped with a BV--operator which rotates the disk boundary
$S^1$.  The algebraic structure on the homology is then a BV--algebra
(Getzler \cite{getz-BV}), where the graded commutative product and the
graded Lie bracket are related by the BV--operator.  Physically,
closed string states associated to each disk boundary $S^1$ are
constrained to be the $S^1$--invariant parts, the kernel of the
BV--operator.  This in turn leads to concentrating on the Lie algebra
structure, where two disk boundaries are identified by {\em
twist}-sewing as Zwiebach did \cite{z:csft}.  On the other hand, he
worked at the chain level (`off shell'), discovering an
$L_\infty$--structure. This was important since the multi-variable
operations of the $L_\infty$--structure provided correlators of closed
string field theory.  Similarly for open string theory, the little
interval operad and associahedra are relevant, the homology
corresponding to a graded associative algebra, but the chain level
reveals an $A_\infty$--structure giving the higher order correlators
of open string field theory.

The corresponding operad for the open--closed string theory is the
Swiss-cheese operad (Voronov \cite{Vo}) that combines the little disk
operad with the little interval operad; it was inspired also by
Kontsevich's approach to deformation quantization.  The algebraic
structure at the homology level has been analyzed thoroughly by
Harrelson \cite{erich}.  In contrast, our work in the open--closed
case is at the level of strong homotopy algebra, combining the known
but separate $L_\infty$-- and $A_\infty$--structures.

In our earlier work, we defined 
such a homotopy algebra and called 
it an {\it open--closed homotopy algebra} (OCHA) \cite{ocha}. 
In particular, 
we showed that this description is a homotopy invariant algebraic
structure, ie,
that it transfers well under homotopy equivalences or quasi-isomorphisms. 
Also, we showed that an open--closed homotopy algebra gives us
a general scheme for deformation of
open string structures ($A_\infty$--algebras)
by closed strings ($L_\infty$--algebras).

In this paper, we aim to explain a background for OCHAs, 
the aspect of moduli spaces of Riemann surfaces. 
Also, we present the relation of OCHAs with Merkulov's geometric 
$A_\infty$--structures \cite{mer2,mer1}.

An open--closed homotopy algebra consists of a direct sum of
graded vector spaces $\cH =\cH_c\oplus\cH_o$. It has
an $L_\infty$--structure on $\cH_c$ and reduces to
an $A_\infty$--algebra if we set $\cH_c=0$. 
Moreover, the 
operations that intertwine the two are a generalization of 
the strong homotopy analog of H Cartan's notion 
of a Lie algebra $\mathfrak g$ 
acting on a differential graded algebra E (Cartan
\cite{cartan:notions}, Flato, Gerstenhaber and Voronov \cite{fgv}). 
In \fullref{sec:oc}, we start from discussing 
the moduli space aspects and the associated operad (tree graph) 
structures 
for $A_\infty$--algebras, $L_\infty$--algebras, and then OCHAs, 
together with recalling other descriptions by 
multi-variable operations and coderivation differentials. 
In a more physically oriented paper \cite{pocha}, 
we gave an alternative interpretation in the language 
of homological vector fields on a supermanifold.

One of the key theorems in homotopy algebra is the minimal model
theorem which was first proved for $A_\infty$--algebras by
Kadeishvili \cite{kadei1}. 
The minimal model theorem states the existence of minimal models 
for homotopy algebras 
analogous to Sullivan's minimal models \cite{Sul} 
for differential  graded commutative algebras 
introduced in the context of rational homotopy theory.
In \fullref{sec:MMth} we 
re-state the minimal model theorem 
for our open--closed homotopy algebras.

As suggested by Merkulov, 
his geometric $A_\infty$--structure \cite{mer2} is 
a special example of an OCHA. 
In \fullref{sec:geom}, we present a new formulation of an OCHA in Merkulov's framework.

In \fullref{ssec:defs}, we recall the relation of open--closed 
homotopy algebras to various aspects of deformation theory
and the relevant moduli spaces and in \fullref{sec:6}, return
to the relation to the motivating string theory. 

There is a distinction between the historical grading
used in defining $A_\infty$-- and $L_\infty$--structures
and the more recent one common in the physics literature.
They are related by (de)suspension of the underlying graded vector spaces. 
Since we emphasize the  versions in terms of
a single differential of degree one on the relevant `standard construction', 
we will only occassionally refer
to the older version, primarily for ungraded strictly 
associative or Lie algebras or  strict differential graded algebras. 
The distinction does influence the exposition, 
but the only importance technically is the signs that occur. 
However, the detailed signs are conceptually unimportant 
(although crucial in calculations), so we indicate them here 
only with $\pm$, the precise details being in \cite{ocha,pocha}.

We restrict our arguments to the case that the characteristic 
of the field $k$ is zero. We further let $k=\C$ for simplicity.

\section{Strong homotopy algebra}
\label{sec:oc}

\subsection{Topology of based loop spaces}
An open--closed homotopy algebra \cite{ocha} is a strong 
homotopy algebra (or $\infty$--algebra) which combines two typical 
strong homotopy algebras, 
an $A_\infty$--algebra and an $L_\infty$--algebra. 

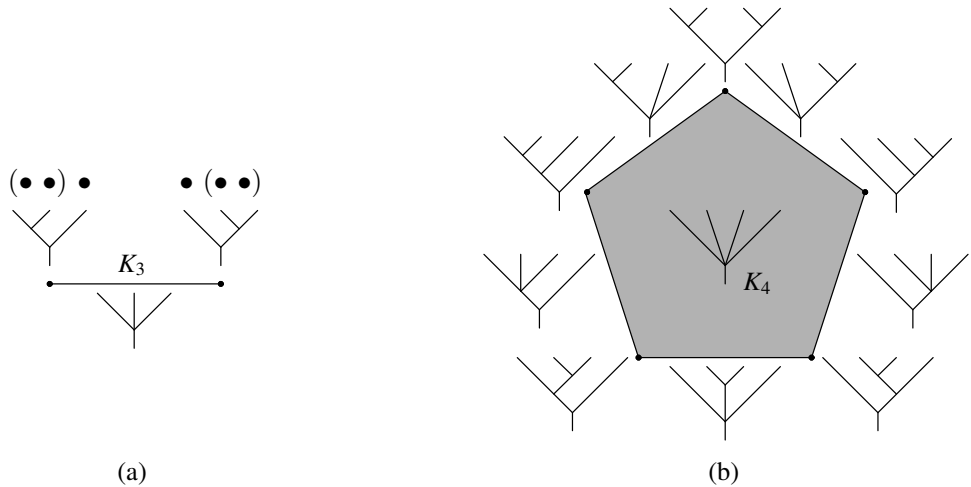
\begin{figure}[ht!]
\labellist\small
\pinlabel (a) [t] at 267 540
\pinlabel (b) [t] at 495 540
\pinlabel $K_3$ [b] at 267 606
\pinlabel $K_4$ at 508 606
\endlabellist
\centering
\includegraphics[width=.97\hsize]{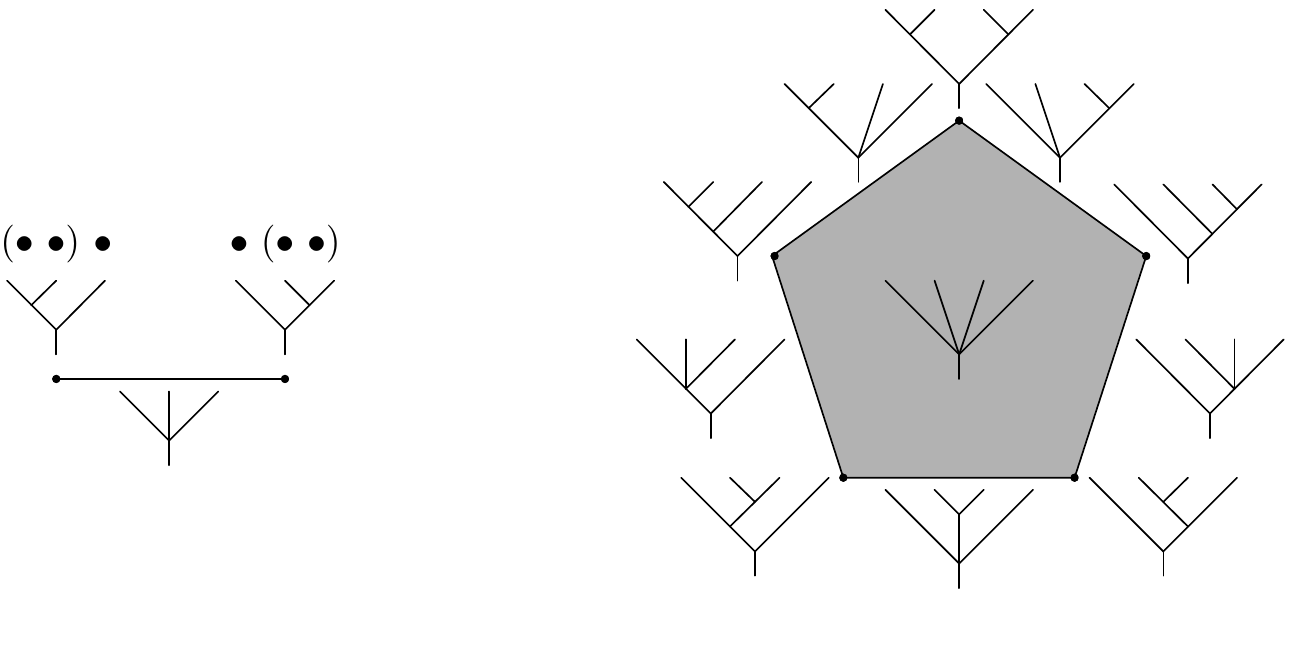} 
 \caption{
(a)\qua An interval as associahedra $K_3$\qquad
(b)\qua A pentagon as\break associahedra $K_4$ 
}
 \label{fig:Ainftysp}
\end{figure}

An $A_\infty$--algebra was introduced \cite{hahI} as a
structure exemplified by the chains on the based loop space $Y:=
\Omega X$ of a topological space $X$ with a base point $x_0\in X.$ A
based loop $x\in Y:= \Omega X$ is a map $x\co[0,1]\to X$ such that
$x(0)=x(1)=x_0$.  The based loop space $Y$ forms a group-like space,
where the product
\begin{equation*}
 m_2 \co  Y\times Y\to Y\ 
\end{equation*}
is given naturally by connecting two based loops as usual. 
The product $m_2$ is not associative but there exists a homotopy 
between $m_2(m_2\times 1)$ and $m_2(1\times m_2)$
described by an interval $K_3$ (\fullref{fig:Ainftysp} (a)) 
\begin{equation*}
 m_3 \co  K_3\times Y\times Y\times Y\lgraw Y\ .
\end{equation*}
In a similar way, we can consider possible operations of 
$(Y)^{\times 4}\to Y$ constructed from 
$m_2$ and $m_3$. 
These connect to form a map on the boundary of a pentagon, 
which can be extended to a pentagon $K_4$ 
(\fullref{fig:Ainftysp} (b)), 
providing a  higher homotopy operation: 
\begin{equation*}
 m_4 \co  K_4\times Y^{\times 4} \lgraw Y\ .
\end{equation*}

\vspace{-5pt}

\vspace{-5pt}
Repeating this procedure leads to higher dimensional
polytopes $K_n$, of dimension $(n-2)$ \cite{hahI}, 
now called {\it associahedra} since 
the vertices correspond to all ways of associating a 
string of $n$ letters. For $Y=\Omega X$, we have higher homotopies 
\begin{equation*}
 m_n \co  K_n\times (Y)^{\times n}\lgraw Y 
\end{equation*}
extending maps on the boundary of $K_n$ defined by compositions of the $m_k$ for $k<n$.
Then, a topological space $Y$ equipped with 
the structures $\{m_n, K_n\}_{n\ge 2}$ as above 
is called an $A_\infty$--space.

\subsection{Compactification of moduli spaces of disks with 
boundary punctures}
\label{ssec:diskmoduli}

\begin{figure}[ht!]
\labellist\small
\pinlabel (a) at 224 590
\pinlabel (b) at 424 590
\pinlabel $\simeq$ at 422 638
\pinlabel $\infty$ [t] at 225 613
\pinlabel $\infty$ [t] at 367 613
\pinlabel 1 [tr] at 194 630
\pinlabel 0 [tl] at 257 630
\endlabellist
\centering
\includegraphics[width=.97\hsize]{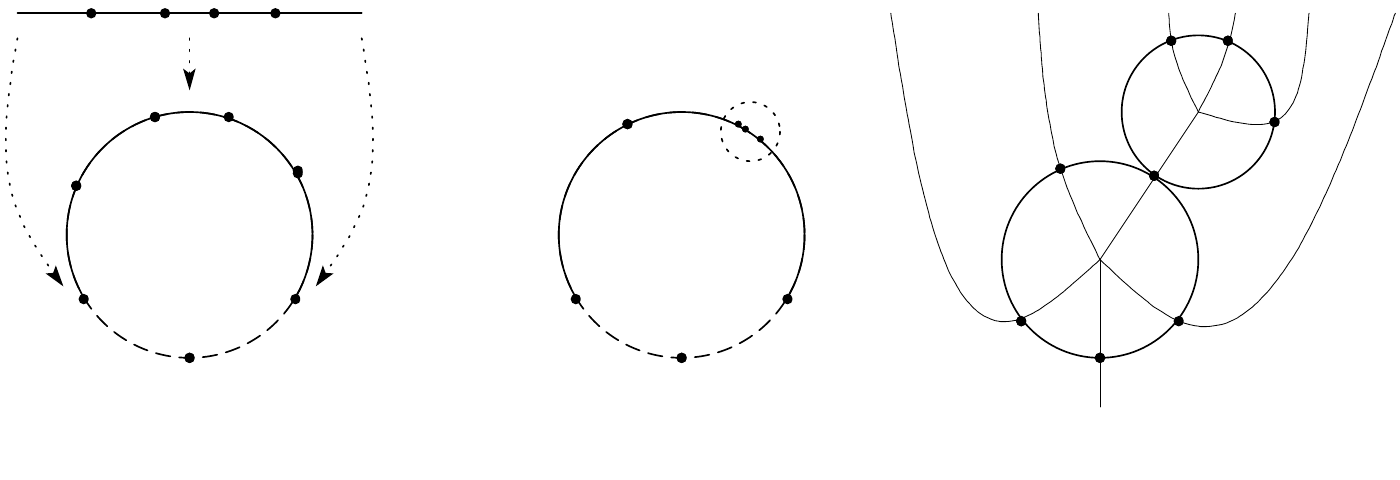}
 \caption{
(a)\qua The identification of the interval with $(n-2)$ points on it 
with the boundary of the disk with $(n+1)$ points on the boundary\newline
(b)\qua The correspondence of the compactification of 
the moduli spaces with tree graphs; the case of a boundary component 
of the compactified moduli space $\cM_7$
}
 \label{fig:comp1}
\end{figure}
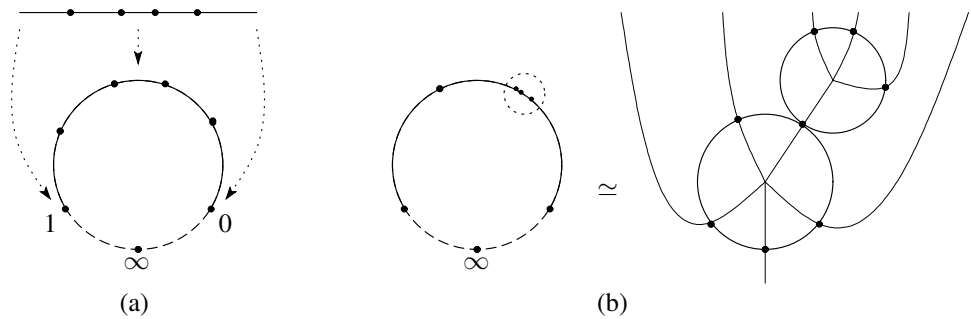

Although it was not noticed for many years, the associahedra $K_n$ can 
be obtained as the moduli space of the real compactification 
of the configuration space of $(n-2)$ distinct points in an interval or 
to a real compactification $\cMub_{n+1}$ of the moduli space 
$\cM_{n+1}$ of a disk with $(n+1)$ points on the boundary 
(\fullref{fig:comp1} (a)). 

This compactification can be related directly to
the planar tree operad (\fullref{fig:comp1} (b)).
$\cM_{n+1}$ is described as the configuration space of 
$(n+1)$--punctures on $S^1\ \sim \R \cup \{\infty\}$ divided by 
conformal transformations. 
In the case in which the Riemann surface is the disk, 
the conformal transformations form $SL(2,\R)$, 
whose element $g\in SL(2,\R)$ acts on $\R\cup\{\infty\}$ as 
\begin{equation}\label{sl2z-transf}
 g(x) =\frac{ax+b}{cx+d}\ ,\quad x\in (\R\cup\{\infty\})\ ,\quad 
 g:=\bp  a & b \\ c & d  \ep \in SL(2,\R)\ .
\end{equation}
This degree of freedom can be killed by 
fixing three points on the boundary. Usually we set the three points 
at $0$, $1$ and $\infty$. We take the point $\infty$ as the `root edge'.

Then, the interval is identified with the arc 
between $0$ and $1$ 
as in \fullref{fig:comp1} (a). Thus, we obtain: 
\begin{equation}\label{cM_n+1}
 \cM_{n+1} = \{ (t_2,\dots, t_{n-1})\ |\ 
 0 < t_2 < t_3 < \cdots < t_{n-1}< 1 \}\ .
\end{equation}
The real compactifications $\cMub_{n+1}$ 
of Axelrod and Singer \cite{as:csII} 
(the real analog of the Fulton--MacPherson compactification \cite{fulmacph}), 
that is, the compactifications of $\cM_{n+1}$ 
with real codimension one boundaries, 
are in fact combinatorially homeomorphic to 
the Stasheff associahedron $K_n$. 
For instance, we rather obviously have:

$\circ$\ for $n=2$, 
$\cM_{2+1}\simeq\{ pt \}\simeq\cMb_{2+1}\simeq K_2$, 

$\circ$\ for $n=3$, $\cM_{3+1}\simeq \{t_2\ |\ 0<t_2 < 1\}$ 
and $\cMb_{3+1}\simeq K_3\simeq$ the closed interval.

For $n>3$, the particulars of the  compactification process
account for the compactification being combinatorially 
homeomorphic to $K_4$ rather than to the closed simplex.

\subsection{Tree formulation}
\label{ssec:Ainftytree}

\begin{figure}[ht!]\small
\centerline{\raise 5pt \hbox{
\labellist
\pinlabel 1 [b] at 95 735
\pinlabel 2 [b] at 114 735
\pinlabel $i$ [b] at 165 735
\pinlabel $k$ [b] at 211 735
\endlabellist 
\includegraphics[scale=.85]{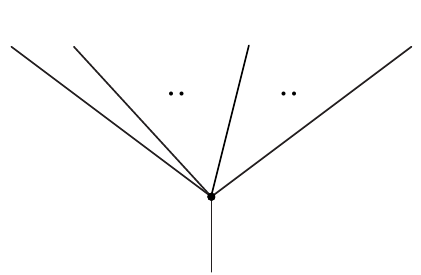}}
\qua \raise 20pt \hbox{$\bullet_i$}\qua 
\raise 5pt \hbox{
\labellist
\pinlabel 1 [b] at 96 735
\pinlabel 2 [b] at 117 735
\pinlabel $l$ [b] at 196 735
\endlabellist 
\includegraphics[scale=.85]{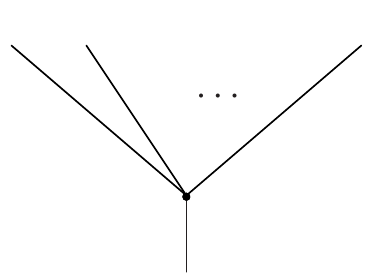}}
\qua \raise 20pt \hbox{=}\qua
\labellist
\pinlabel 1 [b] at 95 735
\pinlabel $i$ [b] at 154 735
\pinlabel $j$ [b] at 196 735 
\pinlabel $n$ [b] at 239 735
\endlabellist 
\includegraphics[scale=.85]{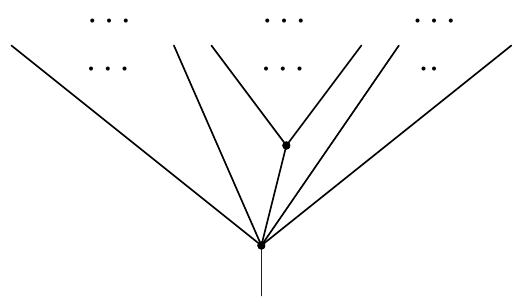}}
\caption{The grafting $M_k\bullet_i M_l$ of the $l$--corolla 
$M_l$ to the $i$-th leaf of $k$--corolla $M_k$, 
where $j=i+l-1$ and $n=k+l-1$}
 \label{fig:grafting}
\end{figure}
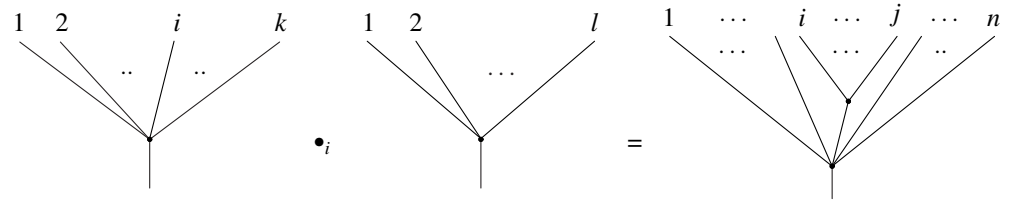

There are some advantages to indexing the maps $m_k$ and 
their compositions by planar rooted trees (as originally suggested by Frank Adams around 1960, when trees would have had to
be inserted in manuscripts by hand); 
e.g. $m_k$ will correspond to the {\em corolla} $M_k$ 
with $k$ leaves all attached directly to the root. 
The composite $m_k\bullet_i m_l$ then corresponds to grafting the root of
$M_l$ to the $i$-th leaf of $M_k$, reading from left to right 
(see \fullref{fig:grafting}). 
(Thus $\bullet_i$ is a precise analog of Gerstenhaber's $\circ_i$, 
although the correspondence was not observed for a couple of decades.)
This is the essence of the planar rooted tree operad \cite{MSS:book}.

Multilinear maps compose in just this way, 
so relations \eqref{Ainfty} can be phrased 
as saying we have a map from planar rooted trees
to multilinear maps respecting the $\bullet_i$ `products', 
the essence of a map of operads \cite{MSS:book}. 
This was originally observed 
in terms of the vector spaces of chains on a based loop space, 
but abstracted as follows:
Let $\cA_\infty(n)$, $n\ge 1$, 
be a graded vector space 
spanned by planar rooted trees with $n$ leaves 
with identity $e\in \cA_\infty(1)$. 
For a planar rooted tree $T\in\cA_\infty(n)$, 
its grading is introduced as 
the number of the vertices contained in $T$, 
which we denote by $v(T)$. 
A tree $T\in \cA_\infty(n)$, $n\ge 2$, with $v(T)=1$ is 
the corolla $M_n$. 
Any tree $T$ with $v(T)=2$ is obtained by the grafting of 
two corollas as in \fullref{fig:grafting}. 
Grafting of any two trees is defined in a similar way, 
with an appropriate sign rule, 
and any tree $T$ with $v(T)\ge 2$ can be obtained recursively by 
grafting a corolla to a tree $T'$ with $v(T')=v(T)-1$. 
One can define a differential $d$ of degree one, which acts on 
each corolla as 
\begin{equation}
 d\left(M_n \right)=- \sum_{k,l\ge 2,\ k+l=n+1}\sum_{i=1}^{k}
 M_k\bullet_i M_l
 \label{Ainfty-operad}
\end{equation}
and extends to one on $\cA_\infty:=\oplus_{n\ge 1}\cA_\infty(n)$ by 
the following rule: 
\begin{equation*}
 d(T\bullet_i T')= d(T)\bullet_i T' +(-1)^{v(T)} T\bullet_i d(T')\ . 
\end{equation*}
If we introduce the contraction of internal edges, that is, 
indicate by $T'\raw T$ that $T$ is obtained from $T'$ by 
contracting an internal edge, the differential is equivalently given 
by 
\begin{equation*}
 d(T)= \sum_{T'\to T}\pm T'\ 
\end{equation*}
with an appropriate sign $\pm$. 
Thus, one obtains a dg operad $\cA_\infty$, which is known as 
the {\em $A_\infty$--operad}. 

An algebra $A$ over $\cA_\infty$ is obtained by 
a representation $\phi\co \cA_\infty(k)\to\Hom(A^{\otimes k},A)$, 
\ie, a map $\phi$ compatible with the $\bullet_i$'s and 
also the differentials. 
We denote by $m_k$ the image $\phi(M_k)$ of $M_k$ by $\phi$. 
Then, 
for each corolla we have 
\begin{equation}
\sum_{k+l=n+1}\sum_{i=1}^{k}
 \pm m_k\bullet_i m_l = 0,
\end{equation}
where we now write $m_1$ for $d$.

This then becomes the definition. 
\begin{defn}[{$A_\infty$--algebra
-- strongly homotopy associative algebra -- \cite{hahI}}]
Let $A$ be a $\Z$--graded vector space 
$A=\oplus_{r\in\Z} A^r$ and suppose that 
there exists a collection of degree one multilinear maps 
\begin{equation*}
 \m:=\{m_k \co  A^{\otimes k}\raw A\}_{k\ge 1} \ .
\end{equation*}
$(A,\m)$ is called an {\em  $A_\infty$--algebra} when the multilinear
maps $m_k$ satisfy the following relations 
\begin{equation}
\sum_{k+l=n+1}\sum_{i=1}^{k}
\pm
 m_k(o_1,\dots,o_{i-1},m_l(o_{i},\dots,o_{i+l-1}),
 o_{i+l},\dots,o_n)=0\ 
 \label{Ainfty}
\end{equation}
for $n\ge 1$. 

A {\em weak} or {\em curved} $A_\infty$--algebra consists of a 
collection of degree one multilinear maps 
\begin{equation*}
 \m:=\{m_k \co  A^{\otimes k}\raw A\}_{k\ge 0} 
\end{equation*} 
satisfying the above relations, but for $n\ge 0$ and 
in particular with $k,l\ge 0$.
 \label{defn:Ainfty}
\end{defn}
\begin{rem}
The `weak' version is fairly new, apparently first in papers
 of Getzler and Jones (and Petrack) \cite{gj:cyclic,gjp}, then was 
adopted by Zwiebach in the $L_\infty$--context \cite{z:csft}
and later used to study what physicists 
refer to  as a `background' for string field theory.
The map  
$m_0\co \C\to A$  is regarded 
as an element $m_0(1)\in A$. 
The augmented relation then implies that 
$m_0(1)$ is a cycle, but $m_1 m_1$ need no longer be 0,  rather
$m_1 m_1 = \pm m_2(m_0\otimes 1) \pm m_2(1\otimes m_0)$. 
 \label{rem:sus}
\end{rem}

\begin{rem}
Recall, as mentioned earlier,  that the component maps would 
have $m_k$ of degree $(k-2)$ in the original formulation.
\end{rem}
\begin{defn}[{$A_\infty$--morphism}]
For two $A_\infty$--algebras $(A, \m)$ and $(A',\m')$,
a collection of degree zero (degree preserving) multilinear maps 
\begin{equation*}
 \{f_k\co  A^{\otimes k}\raw A'\}_{k\ge 1} 
\end{equation*}
is called an {\em $A_\infty$--morphism} 
$\{f_k\}_{k\ge 1}\co (A,\m)\to (A',\m')$ 
iff it satisfies the following relations:
\begin{equation}
\begin{split}
& \sum_{1\leq k_1<k_2\dots <k_j=n}
m'_j(f_{k_1}(o_1,.. ,o_{k_1}),
f_{k_2-k_1}(o_{k_1+1}, .. ,o_{k_2}),\\
&\hspace{3in}
\smash{\dots, f_{n-k_{j-1}}
(o_{k_{j-1}+1}, .. ,o_n))}\\
&\qquad=\sum_{k+l=n+1}\sum_{i=1}^{k}
\pm f_k(o_1,\dots,o_{i-1},m_l(o_{i}, .. ,o_{i+l-1}),
o_{i+l},\dots,o_n)\
\end{split}
\label{amorphism}
\end{equation}
for $n\ge 1$. In particular, if $f_1\co A\to A'$ induces an isomorphism 
between the cohomologies $H(A)$ and $H(A')$, the $A_\infty$--morphism 
is called an {\em $A_\infty$--quasi-isomorphism}. 
 \label{defn:Ainftymorp}
\end{defn}
$A_\infty$--quasi-isomorphisms play important roles from 
the homotopy algebraic point of view (see \fullref{sec:MMth}).

\subsection{Coalgebra formulation}
\label{ssec:Ainf-coalg}

The maps $m_k$ can be assembled into a single map, also
denoted $\m$, from the tensor space $T^cA = \oplus_{k\geq 0}
A^{\ott k}$ to
$A$  with the convention that  $A^{\ott 0}=\C$. The grading implied
by having the maps $m_k$ all of degree one is the usual grading
on each $A^{\ott k}.$ We can regard
$T^cA$ as the tensor {\em coalgebra} by defining
\begin{equation*}
 \tri(o_1 \ott\cdots\ott o_n) 
 = \Sigma_{p=0}^n (o_1\ott\cdots\ott o_p)\ott (o_{p+1}\ott\cdots\ott o_n)\ .
\end{equation*}
A map $f\in \Hom(T^cA, T^cA)$
is a {\em graded coderivation} means 
$\tri f = (f\otimes \1 + \1\otimes f)\tri$, 
with the appropriate signs and dual to the definition of a
graded derivation of an algebra. 
Here $\1$ denotes the identity $\1\co A\to A$. 
We then identify
$\Hom(T^cA, A)$ with $\Coder(T^cA)$ by lifting a multilinear map
as a coderivation \cite{jds:intrinsic}.  Analogously to the situation for
derivations, the composition graded  commutator of coderivations is
again a coderivation; this graded  commutator corresponds 
to the {\em Gerstenhaber bracket} on 
$\Hom(T^cA, A)$ \cite{gerst:coh,jds:intrinsic}. 
Notice that this involves a shift in grading 
since Gerstenhaber uses the traditional Hochschild complex grading. 
Thus $\Coder(T^cA)$ is a graded Lie algebra 
and in fact a dg Lie algebra 
with respect to the bar construction differential, which corresponds 
to the Hochschild differential on $\Hom(T^cA, A)$ 
in the case of an associative algebra $(A,m)$ \cite{gerst:coh}. 
Using the bracket, the differential can be written as $[m,\ ]$.

The advantage of this point of view is that the component maps $m_k$
assemble into a single map $\m$ in $\Coder(T^cA)$ 
and relations \eqref{Ainfty} can be summarized by 
\begin{equation*}
 [\m,\m]=0\  {\rm\quad or, equivalently, \quad} D^2=0\ , 
\end{equation*}
where $D=[\m, \ \,]$. 
In fact, $\m\in \Coder(T^c A)$ is an $A_\infty$--algebra structure 
on $A$ iff $[\m,\m]=0$
{\em and} $\m$ has no constant term, $m_0 =0$. 
If $m_0\neq 0$, the structure is a weak $A_\infty$--algebra.
The $A_\infty$--morphism components similarly combine to give a single map
of dg coalgebras $\f \co T^cA \raw T^cA'$, 
$(\f\otimes\f)\tri=\tri\f$. 
In particular, equation \eqref{amorphism}
is equivalent to 
$\f \circ\m = \m'\circ \f$.

\subsection{$L_\infty$--algebras}
\label{ssec:Linfty}

Since an ordinary Lie algebra $\g$ is regarded as ungraded, the
defining bracket is regarded as skew-symmetric. 
If we regard $\g$ as all of degree one,
then the bracket would be graded symmetric.
For dg Lie algebras and $L_\infty$--algebras, we need graded symmetry,
which refers to the usual symmetry with signs determined by the grading. 
The basic relation is 
\begin{equation}
\tau \co x\otimes y \mapsto (-1)^{|x||y|}y\ott x\ .
\label{tau}
\end{equation}
The sign of a permutation of $n$ graded elements, is defined by
\begin{equation}\label{gcomm}
 \sigma(c_1,\dots .\, ,c_n)
 = \pm(c_{\sigma(1)},\dots .\, ,c_{\sigma(n)}), 
\end{equation}
where the sign $\pm$ is given by what is called the Koszul sign 
of the permutation. 
\begin{defn}[{Graded symmetry}]
A {\em graded symmetric multilinear map}  of a
graded vector space $V$ to itself is a linear map
$f\co V^{\otimes n}\to V$ such that for any $c_i\in V$, $1\le i\le n$, and
any $\sigma\in\S_n$ (the permutation group of $n$ elements), the relation
\begin{equation*}
 f(c_1,\dots .\, ,c_n)
 = \pm f(c_{\sigma(1)},\dots .\, ,c_{\sigma(n)})
\end{equation*}
holds, where $\pm$ is the Koszul sign above. 

The {\em graded symmetric coalgebra} $C(V)$ on a graded vector space $V$ is 
defined as the {\em sub}coalgebra $C(V)\subset T^c V$ 
consisting of the graded symmetric elements in each $V^{\otimes n}$. 
\end{defn}
\begin{defn} By a $(k,l)${\em --unshuffle} of
$c_1,\dots,c_n$ with $n=k+l$ 
is meant a permutation $\sigma$ such that for $i<j\leq k$, 
we have
$\sigma(i)< \sigma(j)$ and similarly for $k < i<j\leq k+l$. 
We denote the subset of $(k,l)$--unshuffles in $\S_{k+l}$
by $\S_{k,l}$ and by $\S_{k+l=n},$ the union of the subgroups
$\S_{k,l}$ with $k+l=n$. 
\end{defn}
\begin{defn}[{$L_\infty$--algebra\ (strong homotopy Lie algebra)
\cite{LS}}]
Let $L$ be a graded vector space and suppose that
a collection of degree one graded symmetric linear maps
$\l:=\{l_k\co L^{\otimes k}\raw L\}_{k\ge 1}$ is given. 
$(L,\l)$ is called an {\em  $L_\infty$--algebra} iff
the  maps satisfy the following relations
\begin{equation}
\sum_{\sigma\in\S_{k+l=n}}\pm
l_{1+l}(l_k(c_{\sigma(1)},\dots,c_{\sigma(k)}),
 c_{\sigma(k+1)},\dots,c_{\sigma(n)})=0\
 \label{Linfty}
\end{equation}
for $n\ge 1$, where $\pm$ is the Koszul sign \eqref{gcomm} 
of the permutation $\sigma\in\S_{k+l=n}$. 

A {\em weak $L_\infty$--algebra} consists of 
a collection of degree one graded symmetric linear maps
$\l:=\{l_k\co  L^{\otimes k}\raw L\}_{l\ge 0}$
satisfying the above relations, but for $n\ge 0$ and 
with $k,l\ge 0$.
 \label{defn:Linfty}
\end{defn}
\begin{rem}
The alternate definition in which the summation is 
over all permutations, rather than just unshuffles, requires 
the inclusion of appropriate coefficients involving factorials. 
\end{rem}
\begin{rem}
A dg Lie algebra is expressed as the desuspension of an 
$L_\infty$--algebra $(L,\l)$ where 
$l_1$ and $l_2$ correspond to the differential and the Lie bracket, 
respectively, and higher multilinear maps $l_3,l_4,\dots$ are absent. 
 \label{rem:susL}
\end{rem}
\begin{rem}
For the `weak/curved' version, remarks analogous to those 
for weak $A_\infty$--algebras apply, and similarly for morphisms.
 \label{rem:weakL}
\end{rem}

In a similar way as in the $A_\infty$ case, 
an $L_\infty$--algebra $(L,\l)$ is described as a coderivation 
$\l\co C(L)\to C(L)$ satisfying $(\l)^2=0$. 
Also, for two $L_\infty$--algebras $(L,\l)$ and $(L',\l')$, 
an $L_\infty$--morphism is defined as a coalgebra map 
$\f\co C(L)\to C(L')$, 
where $\f$ consists of graded symmetric 
multilinear maps $f_k\co L^{\otimes k}\to L'$ 
of degree zero with $k\ge 1$, 
satisfying $\l'\circ\f = \f\circ\l$.

\begin{figure}[ht!]\small
\begin{equation*}
\begin{minipage}[c]{50mm}{
\labellist\small
\pinlabel 1 [b] at 95 735
\pinlabel 2 [b] at 117 735
\pinlabel 3 [b] at 138 735
\pinlabel $k$ [b] at 212 735
\endlabellist
\includegraphics{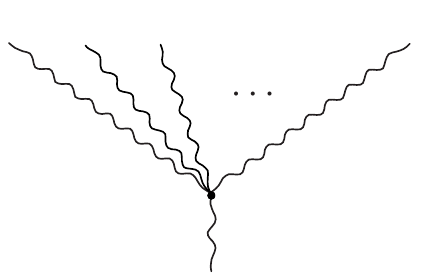}}
\end{minipage}
=
\begin{minipage}[c]{50mm}{
\labellist\small
\pinlabel $\sigma(1)$ [b] at 120 735
\pinlabel $\sigma(2)$ [b] at 143 735
\pinlabel $\sigma(3)$ [b] at 164 735
\pinlabel $\sigma(k)$ [b] at 235 735
\endlabellist
\includegraphics{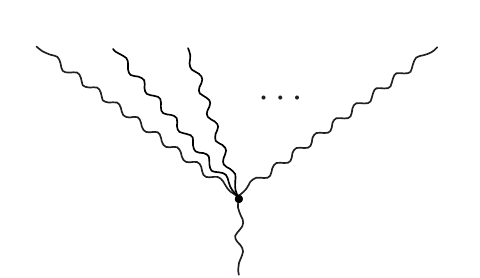}}
\end{minipage}
\end{equation*}
 \caption{Nonplanar $k$--corolla $L_k$ corresponding to $l_k$. 
Since edges are non-planar, it is symmetric with respect to 
the permutation of the edges. }
 \label{fig:l}
\end{figure}
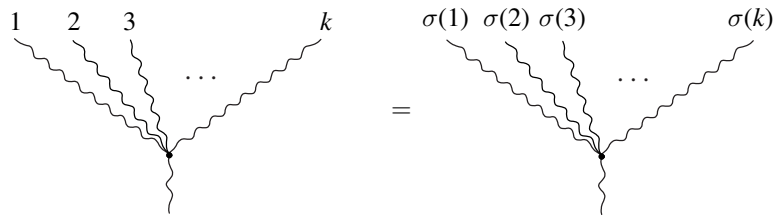

The tree operad description of $L_\infty$--algebras uses
non-planar rooted trees with leaves numbered $1,2,\dots$ arbitrarily 
\cite{MSS:book}. 
Namely, 
a non-planar rooted tree can be expressed as a planar rooted tree 
but with arbitrary ordered labels for the leaves. 
In particular, corollas obtained by permuting the labels are 
identified (\fullref{fig:l}).  
Let $\cL_\infty(n)$, $n\ge 1$, be a graded vector space generated by 
those non-planar rooted trees of $n$ leaves. 
For a tree $T\in\cL_\infty(n)$, a permutation $\sigma\in\S_n$ 
of the labels for leaves generates a different tree in general, 
but sometimes the same one because of the symmetry 
of the corollas above. 
The grafting, $\circ_i$, to the  leaf labelled $i$ is defined  as in the 
planar case in \fullref{ssec:Ainftytree}, 
and any non-planar rooted tree is obtained by 
grafting corollas $\{L_k\}_{k\ge 2}$ recursively, as in the planar case, 
together with the permutations of the labels for the leaves. 
A degree one differential 
$d\co  \cL_\infty(n)\raw\cL_\infty(n)$ is given in a similar way; 
for $T'\raw T$ indicating that $T$ is obtained from $T'$ 
by the contraction of an internal edge, 
\begin{equation*}
 d(T)=\sum_{T'\raw T}\pm T'\ ,
\end{equation*}
and $d(T\circ_i T')= d(T)\circ_i T' +(-1)^{v(T)} T\circ_i d(T')$ 
again holds. 
Thus, $\cL_\infty:=\oplus_{n\ge 1}\cL_\infty(n)$ forms a dg operad, 
called the {\em $L_\infty$--operad}. 

An algebra $L$ over $\cL_\infty$ obtained by a map 
$\phi\co \cL_\infty(k)\raw \Hom(L^{\otimes k},L)$ then forms an 
$L_\infty$--algebra $(L,\l)$.

 \subsection{Compactification of moduli spaces of spheres 
with punctures}

As $A_\infty$--algebras can be described in terms 
of compactifications of moduli spaces of configurations 
of points on an interval, so, with some additional subtlety,
 $L_\infty$--algebras can be described 
in terms of compactifications of moduli spaces 
of configurations of points on a Riemann sphere. 
The compactification corresponding to an $L_\infty$--structure 
is the {\em real} compactification $\cMub_{0,n}$ of the 
moduli spaces $\cM_{0,n}$ of spheres with $n$ punctures 
(\cite{ksv1}, see also \cite{z:csft}). 
Here we use underbar in order to distinguish it from 
the complex compactification by Deligne--Knudsen--Mumford 
which is more familiar and often denoted by $\cMb_{0,n}$. 
Also, we attach the lower index $0$ indicating genus zero, 
in order to distinguish the real compactification of the moduli spaces 
of punctured spheres from that of punctured disks 
in \fullref{ssec:diskmoduli}.

The moduli space $\cM_{0,n}$ is defined as 
the configuration space of $n$ points on 
a sphere $\simeq$ $\C\cup\{\infty\}$ modulo the $SL(2,\C)$ action 
\begin{equation*}
 w'(w)=\frac{aw+b}{cw+d}\ ,\quad w\in\C\cup\{\infty\}\ ,\quad 
\bp a & b\\ c & d\ep\in SL(2,\C).
\end{equation*}
This $SL(2,\C)$ action allows us to fix
three points; usually $0, 1$ and $\infty$. 

For $n=3$, the moduli space $\cM_{0,2+1}$ is a point, so is 
its real compactification $\cMub_{0,2+1}\simeq \{pt\}$. 
For $n=4$, one has 
$$\cM_{0,4}\ \simeq
(\C\cup\infty)-\{0,1,\infty\}, 
$$ 
which is the configuration space of four points 
$0,1,w,\infty$ with the subtraction of the `diagonal'. 
The real compactification of $\cM_{0,4}$ 
looks as in \fullref{fig:pig}: 
\begin{figure}[ht!]\small
\centerline{
\begin{minipage}[c]{55mm}{
\labellist\small
\pinlabel 0 [tl] at 135 672
\pinlabel 1 [tl] at 193 672
\pinlabel $\infty$ [tl] <-2pt,0pt> at 164 698
\pinlabel $B_0$ [br] at 130 681
\pinlabel $B_1$ [bl] at 199 681
\pinlabel $B_\infty$ [br] at 157 707
\endlabellist
\includegraphics{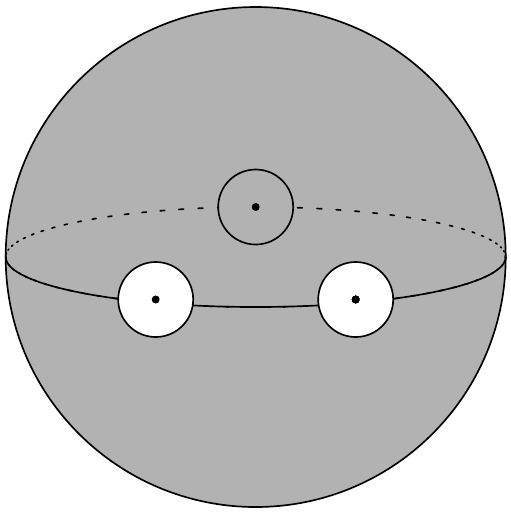}}
\end{minipage}
\ $\simeq$\ \ 
\begin{minipage}[c]{65mm}{
\labellist\small
\pinlabel $B_0$ [bl] at 150 723
\pinlabel $B_1$ [bl] at 215 722
\pinlabel $B_\infty$ [tl] at 239 659
\endlabellist
\includegraphics{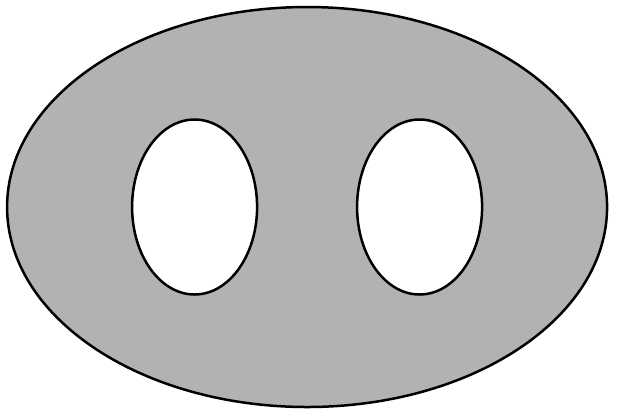}}
\end{minipage}}
 \caption{The real compactification $\cMub_{0,3+1}$ of $\cM_{0,3+1}$}
 \label{fig:pig}
\end{figure}
$\cMub_{0,4}$ has $\codim_\R=1$ boundaries 
$B_0$, $B_1$, $B_\infty$. 
If we associate points $0,1,w$ to $x,y,z$ and 
$\infty$ to the root edge, 
we get the correspondence: 
\renewcommand{\arraystretch}{1.5}
\begin{equation*}
 \begin{array}{ccc}
 B_0 & \lraw & \pm [[x,z],y] \\
 B_1 & \lraw & \pm [[y,z],x] \\
 B_\infty & \lraw & \pm [[x,y],z] \ .
 \end{array}
\end{equation*}
Inspired by closed string field theory, this can be seen 
in terms of `grafting' tubular neighborhoods of trees 
with freedom of a full $S^1$ of rotations of the boundaries 
which are to be identified: \\
\centerline{\begin{minipage}[c]{40mm}{
\labellist\small
\pinlabel $x$ [b] at 152 696
\pinlabel $y$ [b] <0pt,-2pt> at 208 696
\pinlabel $z$ [b] <2pt,0pt> at 235 661
\pinlabel $\infty$ [t] <0pt,-2pt> at 207 594
\endlabellist
\includegraphics{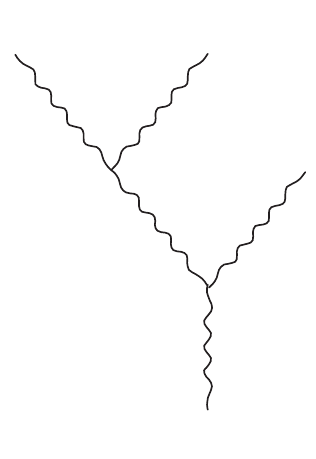}}
\end{minipage}
$\lraw$\ \ 
\begin{minipage}[c]{60mm}{
\labellist\small
\pinlabel $x$ [b] at 164 696
\pinlabel $y$ [b] <0pt,-2pt> at 216 696
\pinlabel $z$ [b] at 239 620
\pinlabel $\infty$ [t] <0pt,-2pt> at 215 563
\pinlabel $S^1$ [r] at 178 628
\endlabellist
\includegraphics{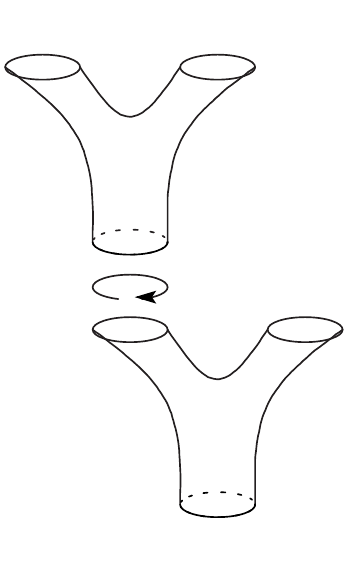}}
\end{minipage}}
Now consider the relative homology groups of the compactified moduli spaces 
modulo those on the lower dimensional strata. These give a
version of the $L_\infty$--operad. 
Corresponding to the relation 
$\partial(\cMub_{0,4})=B_0\coprod B_1\coprod B_\infty$, 
we obtain: 
\begin{equation*}
 d(l_3)(x,y,z)=[[x,y],z]\pm [[y,z],x]\pm [[z,x],y]\ .
\end{equation*}
Notice that $\cMub_{0,n}$ is not contractible for $n\geq 4$. 
In general, $\cMub_{0,n}$ is a manifold with corners 
(as were the associahedra) of 
real dimension $2n-6$, but the strata are not generaly cells,
as they were for the associahedra.  Thus, to define the
dg $L_\infty$--operad, we use the homology of strata relative to boundary 
\cite{ksv1}.
On the other hand, if we are concerned 
only with the corresponding homology operad, we need only use the
little disks operad and Fred's configuration space calculations \cite{cohen}.

 \subsection{Open--closed homotopy algebra (OCHA)}
\label{ssec:ocha}

For our open--closed homotopy algebra, we consider a graded vector space
$\cH=\cH_c\oplus\cH_o$ in which $\cH_c$ will be an $L_\infty$--algebra 
and $\cH_o$, an $A_\infty$--algebra. 

An OCHA is inspired by the compactification of the moduli spaces of 
punctured Riemann surfaces (Riemann surfaces with marked points) 
or the decomposition of the moduli spaces  
as is done in constructing string field theory. 
More precisely, an OCHA should be an algebra over the DG operad 
of chains of the compactified moduli spaces of the corresponding 
Riemann surfaces. 
In this paper, we first present the DG--operad which we call 
the open--closed operad $\cOC_\infty$. 
An OCHA obtained as a representation of 
$\cOC_\infty$ has various interesting structures also 
from purely algebraic points of view. 
In particular, an OCHA can be viewed as a generalization of 
various known algebras. 
We shall discuss this after presenting the definition of an OCHA. 
Before giving the explicit definition in terms of
`algebraic' formulas, we look at the tree description.

 \subsection{The tree description}
\label{ssec:octree}

We associated the $k$--corolla $M_k$ of planar rooted trees to 
the multilinear map $m_k$ of an $A_\infty$--algebra,   
and the $k$--corolla $L_k$ of non-planar rooted trees 
to the graded symmetric multilinear map $l_k$ of an $L_\infty$--algebra. 
For an OCHA $(\cH,\l,\n)$, 
we introduce the $(k,l)$--corolla $N_{k,l}$ 
\begin{equation}
N_{k,l}\ \ =\ \  
\begin{minipage}[c]{45mm}{
\labellist\small
\pinlabel $1$ [b] at 95 735
\pinlabel $1$ [b] at 163 735
\pinlabel $k$ [b] at 142 735
\pinlabel $l$ [b] at 210 735
\endlabellist
\includegraphics{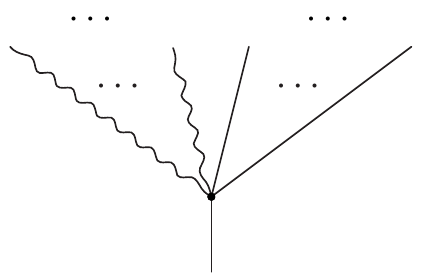}}
\end{minipage}
 \label{n_kl}
\end{equation}
which is defined to be partially symmetric (non-planar); 
only symmetric with respect to the $k$ leaves. 
We express symmetric leaves as wiggly edges and 
planar (= non-symmeteric) leaves as straight edges as before. 
Let us consider such corollas for $2k+l+1\ge 3$. 
As we shall explain further later, 
this constraint is motivated by the stable moduli space of 
a disk with $k$--punctures interior and $(l+1)$--punctures 
on the boundary of the disk. 
We also consider non-planar corollas $\{L_k\}_{k\ge 2}$. 
The planar $k$--corolla $M_k$ is already included as $N_{0,k}$. 
Since we have two kinds of edges, we have two kinds of grafting. 
We denote by $\circ_i$ (resp. $\bullet_i$) the grafting 
of a wiggly (resp. straight) root edge to 
an $i$-th wiggly (resp. straight) edge, respectively. 
For these corollas, we have three types of composite; 
in addition to the composite $L_{1+k}\circ_i L_l$ in $\cL_\infty$, 
there is a composite $N_{k,m}\circ_i L_p$ described by 
\begin{equation*}
\begin{minipage}[c]{40mm}{
\labellist\small
\pinlabel $1$ [b] at 95 735
\pinlabel $1$ [b] at 163 735
\pinlabel $k$ [b] at 142 735
\pinlabel $m$ [b] at 210 735
\endlabellist
\includegraphics[scale=.85]{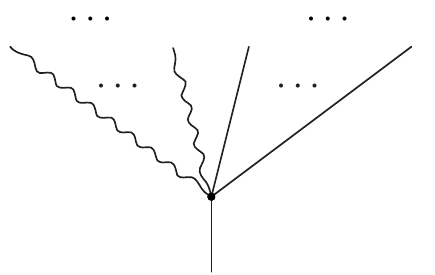}}
\end{minipage}
\circ_i
\begin{minipage}[c]{37mm}{
\labellist\small
\pinlabel $1$ [b] at 94 735
\pinlabel $2$ [b] at 117 735
\pinlabel $3$ [b] at 139 735
\pinlabel $p$ [b] at 209 735
\endlabellist
\includegraphics[scale=.85]{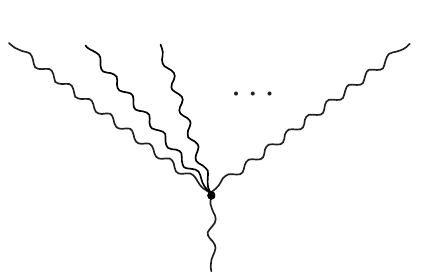}}
\end{minipage}\ =\ 
\begin{minipage}[c]{45mm}{\includegraphics[scale=.85]{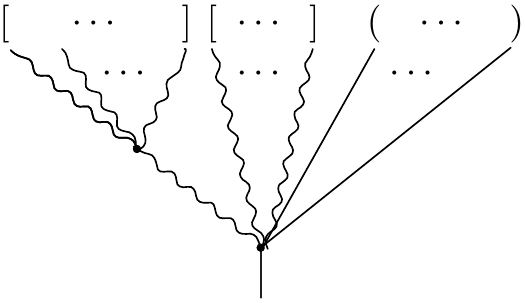}}
\end{minipage}
\end{equation*}
where in the right hand side the labels are given by 
$[i,\dots,i+p-1][1,\dots,i-1,i+p,\dots,p+k-1](1,\dots,m)$, 
and the composite $N_{p,q}\bullet_i N_{r,s}$ 
\begin{equation*}
\begin{minipage}[c]{40mm}{
\labellist\small
\pinlabel $1$ [b] at 95 735
\pinlabel $1$ [b] at 163 735
\pinlabel $p$ [b] at 142 735
\pinlabel $q$ [b] at 210 735
\endlabellist
\includegraphics[scale=.85]{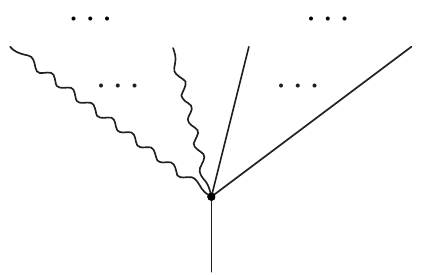}}
\end{minipage}
\bullet_i
\begin{minipage}[c]{37mm}{
\labellist\small
\pinlabel $1$ [b] at 95 735
\pinlabel $1$ [b] at 163 735
\pinlabel $r$ [b] at 142 735
\pinlabel $s$ [b] at 210 735
\endlabellist
\includegraphics[scale=.85]{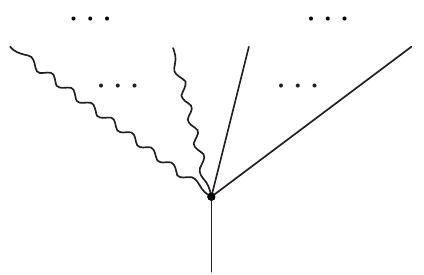}}
\end{minipage}\ =\ 
\begin{minipage}[c]{45mm}{
\includegraphics[scale=.85]{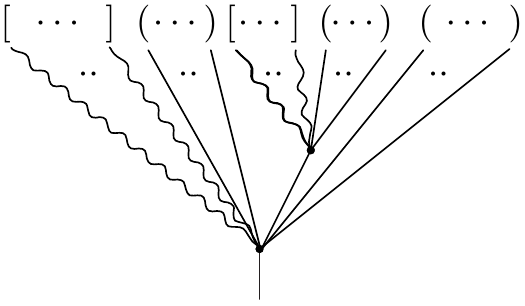}}
\end{minipage}\ 
\end{equation*}
with labels 
$[1,\dots,p](1,\dots,i-1)[p+1,\dots,p+r](i,\dots,i+s-1)(i+s,\dots,q+s-1)$.
To these resulting trees, grafting of a corolla $L_k$ or $N_{k,l}$ 
can be defined in a natural way, and we can repeat this procedure. 
Let us consider tree graphs obtained in this way, 
that is, by grafting the corollas $l_k$ and $n_{k,l}$ recursively, 
together with 
the action of permutations of the labels for closed string leaves. 
Each of them has a wiggly or straight root edge. 
The tree graphs with wiggly root edges, 
with the addition of the identity $e_c\in\cL_\infty(1)$, 
generate $\cL_\infty$ as stated in 
\fullref{ssec:Linfty}. 
On the other hand, 
the tree graphs with both wiggly and straight edges are new. 
\begin{defn}We denote by $\cN_\infty(k;l)$, 
the graded vector space generated by rooted tree graphs with 
$k$ wiggly leaves and $l$ straight leaves. 
In particular, we formally add the identity $e_o$ generating 
$\cN_\infty(0;1)$ and a corolla
$N_{1,0}$ generating $\cN_\infty(1;0)$. 
The tree operad relevant for OCHAs is then 
$\cOC_\infty:=\cL_\infty\oplus\cN_\infty$. 
\end{defn}
In fact, $\cOC_\infty$ is an example of 
a {\em colored} operad \cite{BoVo,MSS:book,laan:03}. 
For each tree $T\in\cOC_\infty$, 
its grading is given by the number of  vertices $v(T)$. 

{}For trees in $\cOC_\infty$, 
let $T'\raw T$ indicate that $T$ is obtained from $T'$ by 
contracting a wiggly or a straight internal edge. 
A degree one differential $d\co  \cOC_\infty\raw\cOC_\infty$ 
is given by 
\begin{equation}\label{diff-ocoperad}
  d(T)= \sum_{T'\raw T}\pm T'\ ,
\end{equation}
so that the following compatibility holds: 
\begin{equation*}
 \begin{split}
 &d(T\circ_i T')=d(T)\circ_i T' + (-1)^{v(T)} T\circ_i d(T'), 
\\
 & d(T\bullet_i T'')=d(T)\bullet_i T'' 
 + (-1)^{v(T)} T\bullet_i d(T'')\ .
 \end{split}
\end{equation*}
Thus, $\cOC_\infty$ forms a dg colored operad.

\subsection{Formal definition of OCHA}
\label{ssec:formal-ocha}

For two $\Z$--graded vector spaces $\cH_c$ and $\cH_o$, 
an open--closed homotopy algebra $(\cH:=\cH_c\oplus\cH_o,\l,\n)$ 
is an algebra over the operad $\cOC_\infty$. 
An algebra $\cH:=\cH_c\oplus\cH_o$ over $\cOC_\infty$ is obtained by 
a representation 
\begin{equation*}
 \phi\co \cL_\infty(k)\to \Hom(\cH_c^{\otimes k},\cH_c) \ ,\qquad 
 \phi\co \cN_\infty(k;l)\to 
 \Hom((\cH_c)^{\otimes k}\otimes (\cH_o)^{\otimes l},\cH_o) \ 
\end{equation*}
which is compatible with respect to the grafting 
$\circ_i$, $\bullet_i$ and the differential $d$. 
Here, regarding elements in 
both $\Hom(\cH_c^{\otimes k},\cH_c)$ and 
$\Hom((\cH_c)^{\otimes k}\otimes (\cH_o)^{\otimes l},\cH_o)$ 
as those in $\Coder(C(\cH_c)\otimes T^c(\cH_o))$, 
the differential on the algebra side is given by 
\begin{equation}\label{diff-oc}
d:=[l_1+n_{0,1},\ \ ], 
\end{equation}
where both $l_1$ and $n_{0,1}$ are the canonical lift of 
differentials $l_1\co \cH_c\to\cH_c$ and $n_{0,1}\co \cH_o\to\cH_o$ 
on the corresponding graded vector spaces. 
On the other hand, 
in addition to the differential $d\co \cL_\infty\to\cL_\infty$ 
defining the $L_\infty$--structure, 
we have the differential $d\co \cN_\infty\to\cN_\infty$ 
\eqref{diff-ocoperad} which acts 
on the corolla $N_{k,l}$ as 
\begin{equation*}
 d(N_{n,m})=\sum_{k+p=n+1}\sum_{i}N_{k,m}\circ_i L_p 
 + \sum_{p+r=n,q+s=m+1}\sum_{i} N_{p,q}\bullet_i N_{r,s}. 
\end{equation*}
By combining this with equation \eqref{diff-oc}, 
one can write down the conditions for an OCHA: 
\begin{equation*}
 \begin{split}
0=& \sum_{\sigma\in\S_{p+r=n}}\pm
n_{1+r,m}\left( 
(l_p\otimes \1_c^{\otimes r}\otimes\1_o^{\otimes m})
(c_{\sigma(I)};o_1,\dots,o_m) \right) \\
& +
\sum_{\substack{\sigma\in\S_{p+r=n}\\i+s+j=m}}
\pm
n_{p,i+1+j}\left( (\1_c^{\otimes p}\otimes \1_o^{\otimes i}\otimes
n_{r,s}\otimes \1_o^{\otimes j} )
(c_{\sigma(I)};o_1,\dots,o_m) \right), 
 \end{split}
\end{equation*}
where $c_1,\dots,c_n$ and $o_1,\dots,o_m$ are homogeneous elements in 
$\cH_c$ and $\cH_o$, respectively, 
and the signs $\pm$ are the Koszul sign \eqref{gcomm} of $\sigma$. 

More explicitly: 
\begin{defn}[{Open--Closed Homotopy Algebra (OCHA) \cite{ocha}}]
An {\em open--closed homotopy algebra (OCHA)} 
$(\cH=\cH_c\oplus \cH_o, \l, \n)$ consists of 
an $L_\infty$--algebra $(\cH_c,\l)$ and a family of maps 
$\n=\{n_{p,q}\co \cH_c^{\ott p}\ott\cH_o^{\ott q} \to\cH_o\}$ 
of degree one 
for $p,q\geq 0$ with the exception of $(p,q)=(0,0)$ 
satisfying the compatibility conditions: 
\begin{equation}
\begin{split}
&0=\sum_{\sigma\in\S_{p+r=n}}\pm
n_{1+r,m}(l_p(c_{\sigma(1)},\dots,c_{\sigma(p)}),
c_{\sigma(p+1)},\dots,c_{\sigma(n)};o_1,\dots,o_m) \\
& \hbox to 3pt{\qquad+\hss}
\sum_{\sigma\in\S_{p+r=n},\ i+s+j=m}
\pm
n_{p,i+1+j}(c_{\sigma(1)},.. ,c_{\sigma(p)};o_1, .. ,o_i, \\
&\hspace*{4.0cm} n_{r,s}(c_{\sigma(p+1)}, .. ,c_{\sigma(n)};o_{i+1}, .. ,o_{i+s}),
o_{i+s+1}, .. ,o_m)  
\end{split}
\end{equation}
for homogeneous elements $c_1,\dots,c_n\in\cH_c$ and $o_1,\dots,o_m\in\cH_o$ 
with the full range $n,m\ge 0$, $(n,m)\ne (0,0)$. 
The signs $\pm$ are given in \cite{ocha}. 

A {\em weak/curved OCHA} consists of 
a weak $L_\infty$--algebra $(\cH_c,\l)$ with a family of maps 
$\n=\{n_{p,q}\co  \cH_c^{\ott p}\ott\cH_o^{\ott q} \to\cH_o\}$, 
of degree one, 
now for $p,q\geq 0$ satisfying the analog of the above relation.
 \label{defn:ocha}
\end{defn}
An open--closed homotopy algebra includes various sub-structures 
or reduces to various simpler structures as particular cases. 
The substructure 
$(\cH_c,\l)$ is by definition an $L_\infty$--algebra and 
$(\cH_o,\{m_k:=n_{0,k}\})$ forms an $A_\infty$--algebra. 
A nontrivial structure obtained as a special case of OCHAs is 
the action of
$\cH_c$ as an $L_\infty$--algebra on $\cH_o$ as a dg vector space. 
Lada and Markl (\cite{lada-markl} Definition 5.1) provide 
the definition of an $L_\infty$--module 
where one can see the structure 
as satisfying the relations for a Lie module `up to homotopy'. 
This is the appropriate strong homotopy version of the action 
of an ordinary Lie algebra $L$ on a vector space $M$, also described 
as $M$ being a module over $L$ or a representation of $L$. 
If we set $n_{p,0}=0$ for all $p\ge 1$, the substructure
$(\cH,\{n_{p,1}\})$ makes $\cH_o$ an $L_\infty$--module over $(\cH_c,\l)$. 
Thus we can also speak of $\cH_o$ as a {\em strong homotopy module} 
over $\cH_c$ or as a {\em strong homotopy representation} 
of $\cH_c$ (cf \cite{jds:bull}). 

On the other hand, 
$n_{1,q}$ with $q\ge 1$ forms a strong homotopy derivation 
\cite{markl:1203} with respect to the 
$A_\infty$--algebra $(\cH_o,\{m_k\}_{k\ge 1})$. 
Moreover, we have the strong 
homotopy version of an algebra $A$ over a Lie algebra $L$, that is, an
action of $L$ by derivations of $A$, 
so that the $L_\infty$--map $L\to \End(A)$
takes values in the Lie sub-algebra $\Der A$.

In his ground breaking ``Notions d'alg\`ebre diff\'erentielle; $\cdots$ ''
\cite{cartan:notions}, 
Henri Cartan formalized several dg algebra notions related to his study of
the deRham cohomology of principal fibre bundles, in particular, that
of a Lie group $G$ acting in (`dans') a differential graded algebra E. 
The action uses only the Lie algebra $\mathfrak g$ of $G$. 
Cartan's action includes both the graded derivation $d$, the Lie derivative
$\theta(X)$ and the inner derivative $i(X)$ 
for $X\in\mathfrak g$. The concept was later reintroduced 
by Flato, Gerstenhaber and Voronov \cite{fgv} 
under the name {\em Leibniz pair}. 

We need only the $\theta(X)$ 
(which we denote $\rho(X)$ since by $\theta$ we denote the image 
by $\rho$ of an element $X\in\g$), 
then the algebraic structure is an example of 
a dg algebra over a dg Lie algebra $\g$. 
Its higher homotopy extension 
a mathematician would construct 
by the usual procedures of strong homotopy algebra 
leads to the following definition 
(see the Appendix by M~Markl in \cite{ocha}): 
\begin{defn}[{$A_\infty$--algebra over an $L_\infty$--algebra}]
Let $L$ be an $L_\infty$--algebra and $A$ an $A_\infty$--algebra 
which as a dg vector space is an sh-L module. 
That $A$ is an {\em $A_\infty$--algebra over $L$} means 
that the module structure map
$\rho\co L\to \End(A)$, regarded as in $\Coder(T^cA)$, extends to
an $L_\infty$--map $L\to \Coder(T^cA)$. 
 \label{defn:AovL}
\end{defn}
An $A_\infty$--algebra over an $L_\infty$--algebra defined as above 
is an OCHA $(\cH,\l,\n)$ with $n_{p,0}=0$ for $p\ge 1$.

Given two OCHAs $(\cH,\l,\n)$ and $(\cH',\l',\n')$, 
an OCHA morphism from $(\cH,\l,\n)$ to $(\cH',\l',\n')$ 
is defined by a collection of degree zero multilinear maps 
$f_{k}\co (\cH_c)^{\otimes k}\to\cH_c'$, $k\ge 1$, 
and 
$f_{k,l}\co (\cH_c)^{\otimes k}\otimes (\cH_o)^{\otimes l}\to\cH_o'$, 
$k,l>0$, $(k,l)\ne (0,0)$, 
satisfying certain conditions \cite{ocha}. 
In particular, $\{f_k\}_{k\ge 1}$ forms an $L_\infty$--morphism 
from $(\cH_c,\l)$ to $(\cH_c',\l')$. 
The notion of {\em OCHA--quasi-isomorphisms} is 
defined as OCHA--morphisms such that both $f_1\co \cH_c\to \cH_c'$ and 
$f_{0,1}\co \cH_o\to\cH_o'$ induce isomorphisms on the cohomologies.

An OCHA $(\cH,\l,\n)$ has a coalgebra description 
in terms of a degree one codifferential constructed from 
$\l$ and $\n$ on the tensor coalgebra of $\cH$ (see \cite{ocha}). 
Hoefel \cite{ed:coder} has recently shown the following:
\begin{thm}[{Hoefel \cite{ed:coder}}]
OCHAs are characterized as being given 
by {\it all} coderivations of degree 1 and square zero
on $\Coder(C(\cH_c)\otimes T^c(\cH_o))$.
\end{thm}
Then, two OCHAs $(\cH,\l,\n)$ and $(\cH',\l',\n')$, 
an OCHA--morphism $\f\co (\cH,\l,\n)\to(\cH',\l',\n')$ 
is described as a coalgebra map 
$\f\co C(\cH_c)\otimes T^c(\cH_o)\to C(\cH_c')\otimes T^c(\cH_o')$ 
satisfying $(\l'+\n')\circ\f=\f\circ(\l+\n)$, 
where $\l+\n$ is the codifferential 
constructed from the OCHA structures $\l$ and $\n$. 

Also, one can describe an OCHA dually in terms of 
a supermanifold, see \cite{pocha}.

 \subsection{Examples of the moduli space description}

An OCHA should be an algebra over the DG operad of relative
chains of the compactified moduli spaces of the corresponding 
punctured Riemann surfaces. 
The strata of the compactified moduli space can be labelled by 
the trees of the $\cOC_\infty$--operad. In this direction, 
Hoefel discusses more carefully the 
details of these structures \cite{ed:geo-ocha}.

Let us consider a moduli space corresponding to the open--closed case. 
For $n_{p,q}$, the corresponding moduli space is that of 
a disk with $p$--punctures in the bulk (interior) 
and $(q+1)$--punctures on the boundary. 
For $p=0$, as we saw, $\{m_q=n_{0,q}\}$ forms an $A_\infty$--structure, 
and the corresponding moduli spaces are the associahedra. 
The moduli space corresponding to the operation 
$\{n_{1,q}\}$ with one closed string input is the same as the 
{\em cyclohedra} $\{W_{q+1}\}$, which is 
the moduli space of configuration space of points on $S^1$ 
modulo rotation discussed by Bott and Taubes (see \cite[page
  241]{MSS:book}
and \cite{Staconf}). 

\begin{figure}[ht!]
\centerline{
\raise20pt\hbox{
\labellist\small
\pinlabel (a) <0pt,-30pt> at 256 529
\pinlabel (b) <120pt,-30pt> at  256 529
\pinlabel $W_2$ [t] at  256 529
\pinlabel $W_2$ [b] at 256 646
\pinlabel $S^1$ [t] at 256 627
\endlabellist
\includegraphics[scale=.75]{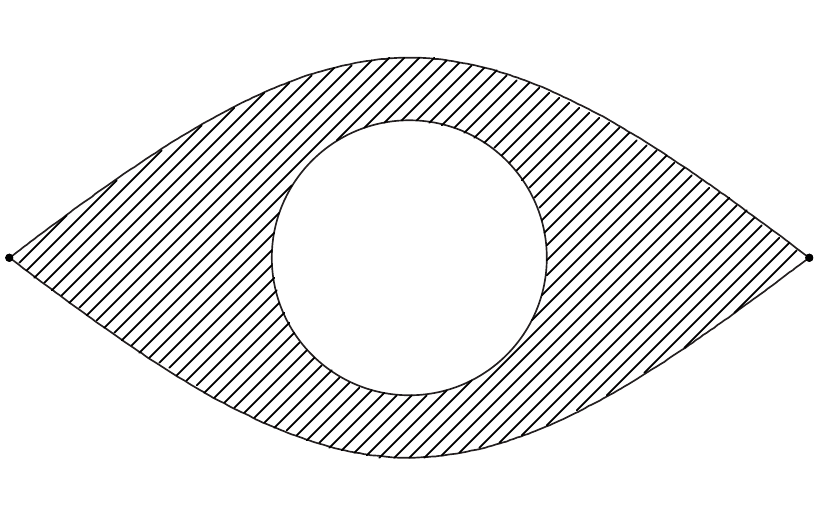}\hspace*{1cm}}
\includegraphics[scale=.75]{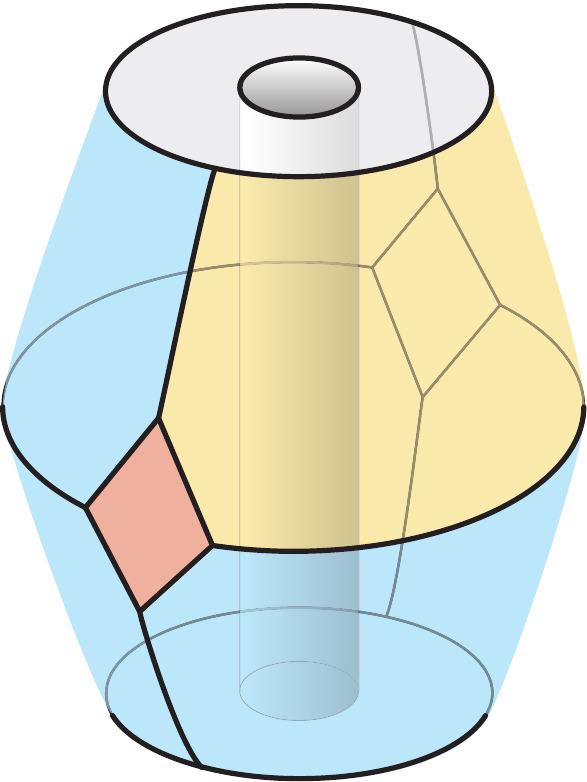}}
\caption{(a)\qua The Eye\quad (b)\qua The solid torus}
 \label{fig:n_2q}
\end{figure}
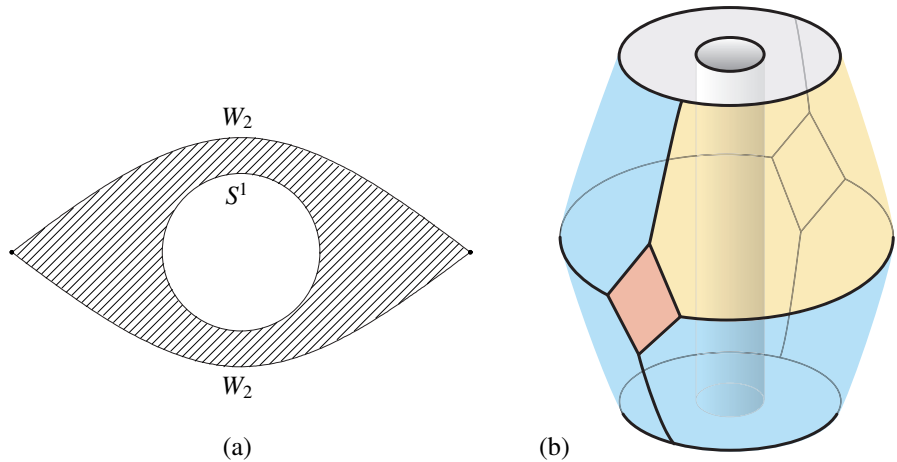

The cyclohedra $\{W_n\}$ are contractible polytopes. 
However, the moduli spaces corresponding to $n_{p,q}$ with 
$p\ge 2$ are not contractible in general. 
Let us consider the moduli space corresponding to $n_{2,q}$: 
the disk with two two interior punctures (= closed strings). 
For $q=0$, the moduli space is described as in \fullref{fig:n_2q} (a), 
which is what is called `The Eye' in the paper 
on deformation quantization by Kontsevich \cite{kont:defquant-pub}. 
For $n_{2,1}$, the compactified moduli space is 
topologically a solid torus as in \fullref{fig:n_2q} (b) 
(this figure was made by S~Devadoss \cite{Devadoss}).

Recall that the facets (codim one faces) of the associahedra
are products of associahedra.  For cyclohedra, the facets
are products of cyclohedra and associahedra. 
The analog holds for the open--closed compactified moduli spaces 
(which currently are nameless), 
although they are not polytopes.

\subsection{Cyclic structure}

We can also consider an additional {\em cyclic structure} 
on open--closed homotopy algebras. See \cite{ocha,pocha} for the definition. 
The cyclic structure can be defined 
in terms of  {\em symplectic} inner products.
These inner products are 
 essential to the description of the Lagrangians 
appearing in string field theory (see \cite{thesis}). 
The string theory motivation for this additional structure is 
that punctures on the boundary
of the disk inherit a cyclic order from the orientation of the disk
and the operations are to respect this cyclic structure, just as the
$L_\infty$--structure reflects the symmetry of the punctures 
in the interior of the disk or on the sphere.

Let us explain briefly the cyclicity in the case of 
$A_\infty$--algebras. 
In terms of trees, the distinction between the root and the leaves can be
absorbed by regarding these edges as cyclically ordered.

{}From the viewpoint of the moduli spaces $\cM_{n+1}$ of punctured disks, 
representing $\cM_{n+1}$ as in equation \eqref{cM_n+1}, 
the cyclic action is an automorphism $g\co  \cM_{n+1}\to\cM_{n+1}$, 
where $g$ is an $SL(2,\R)$ transformation \eqref{sl2z-transf} 
such that 
\begin{equation*}
 \begin{split}
 (\infty,0=t_1,t_2,\dots,t_{n-1},t_n=1)\mapsto 
& (g(\infty),g(0),g(t_2),\dots,g(t_{n-1}),g(1)) \\
& \ \ \ \ = (0,t_2,\dots,t_{n-1},1,\infty). 
 \end{split}
\end{equation*}
One can compactify $\cM_{n+1}$ so that the cyclic action extends to 
the one on $\cMb_{n+1}$. 
In this way, one can consider a cyclic action on the associahedra. 
This cyclic action for the associahedra is discussed in \cite{GeK1}. 
Thus, from the viewpoint of Riemann surfaces, \ie, string theory, 
taking the cyclic action into account for the associahedra is 
very natural. 

This can also be seen visually from the planar trees associated
to disks with points on the boundary, cf \fullref{fig:comp1} (b).
Correspondingly, for an $A_\infty$--algebra $(A,\m)$, a cyclic structure is 
defined by a (nondegenerate) inner product 
$\omega\co A\otimes A\to\C$ of fixed integer degree satisfying 
\begin{equation*}
 \omega(m_n(o_1,\dots,o_n),o_{n+1})=\pm 
 \omega(m_n(o_2,\dots,o_n,o_{n+1}),o_1) 
\end{equation*}
for any homogeneous elements $o_1,\dots,o_{n+1}\in A$ 
(see \cite{MSS:book}). 

In a similar way using a nondegenerate inner product, 
a cyclic structure is defined for an $L_\infty$--algebra 
and then for an OCHA \cite{ocha,pocha}.

\section{The minimal model theorem}
\label{sec:MMth}

Homotopy algebras are designed to have homotopy invariant properties. 
A key and useful theorem in homotopy algebras is then 
the minimal model theorem, proved by Kadeishvili 
for $A_\infty$--algebras \cite{kadei1}. 
For an $A_\infty$--algebra $(A,\m)$, the minimal model theorem states 
that there exists another $A_\infty$--algebra $(H(A),\m')$ 
on the cohomologies of $(A,m_1)$ and an $A_\infty$--quasi-isomorphism 
from $(H(A),\m')$ to $(A,\m)$. 
Since $\m'$ is an $A_\infty$--structure on the cohomology $H(A)$, 
the differential $m_1'$ is trivial; such an $A_\infty$--algebra is called 
{\em minimal}.

The minimal model theorem holds also for an OCHA, which implies 
that 
an OCHA is also appropriately  called a homotopy algebra: 
\begin{defn}[{Minimal open--closed homotopy algebra}]
An OCHA $(\cH=\cH_c\oplus\cH_o,\l,\n)$ 
is called {\it minimal} if 
$l_1=0$ on $\cH_c$ and $n_{0,1}=0$ on $\cH_o$. 
 \label{defn:minimalalg}
\end{defn}
\begin{thm}[{Minimal model theorem for open--closed homotopy algebras}]
{}For a given OCHA $(\cH,\l,\n)$, 
there exists a minimal OCHA $(H(\cH),\l',\n')$ and 
an OCHA--quasi-isomorphism $\f\co  (H(\cH),\l',\n')\to (\cH,\l,\n)$. 
 \label{thm:minimal}
\end{thm}
(See \fullref{ssec:formal-ocha} for the definition of 
an OCHA--quasi-isomorphism. )

Various stronger versions of this minimal model theorem hold for OCHAs
\cite{ocha}, as for $A_\infty$--algebras, $L_\infty$--algebras, etc.
One of them is the {\em homological perturbation theory} developed in
particular on the homology of a differential graded algebra
\cite{gugen,hueb-kadei,GS,gugen-lambe,GLS:chen,GLS} (see \cite{mer}
for an application) and of a dg Lie algebra \cite{jh-jds}.  Another
one is the {\em decomposition theorem} (see \cite{KaTe,thesis}).  As
for classical $A_\infty$, $L_\infty$, etc cases, these theorems imply
that OCHA--quasi-isomorphisms and in particular the one in
\fullref{thm:minimal} in fact give homotopy equivalence between OCHAs.
This further implies the uniqueness of a minimal model for an OCHA
$(\cH,\l,\n)$; a minimal OCHA $H(\cH)$ is unique up to an isomorphism
on $H(\cH)$.

\section{Geometric construction of OCHAs and Merkulov's structures}
\label{sec:geom}

As mentioned, OCHAs admit a geometric expression 
in terms of supermanifolds as given in \cite{pocha}. 
Here, instead of that, we give a partially supermanifold description 
in which an OCHA can be viewed as a `geometric' weak 
$A_\infty$--structure. 
This description is inspired by Merkulov's geometric $A_\infty$-- (and 
$C_\infty$)--structures discussed 
as a generalization of Frobenius structures \cite{mer2}. 

\begin{defn}[{Merkulov \cite{mer1,mer2} -- paraphrased}]
A (Merkulov) {\em geometric} $A_\infty$--structure on a graded manifold 
$ M$ with its tangent bundle $TM$ is a
collection of maps for $n\geq 1$:
\begin{itemize}
 \item[(i)] 
 $\mu_n\co \otimes^n_{{\mathcal O}_M} {\mathcal T}_M \to {\mathcal T}_M$,   
 \item[(ii)] where $\mu_1:= [\nu,\ ]$ is defined 
in terms of  an element $\nu\in {\mathcal T}_M$ 
such that $[\nu,\nu]=0$, 
\end{itemize}
\vspace{-10pt}
making the sheaf of sections ${\mathcal T}_M$  of
$TM$ into a sheaf of $A_\infty$--algebras.
\end{defn}
Condition (ii) means that $\nu$ is a homological vector field 
on ${\mathcal T}_M$, cf the dual supermanifold description of 
an $A_\infty$--structure.

The Merkulov geometric $A_\infty$--structure 
can be obtained as a special case of an OCHA $(\cH:=\cH_c\oplus\cH_o, \l,\n)$ 
in which the $\Z$--graded supermanifold is an $L_\infty$--algebra $\cH_c$.
With an eye toward the relevant deformation theory, 
we use the formal graded commutative power series ring 
denoted by $\C[[\psi]]$.
More precisely, denote by $\{\eb_{i}\}$ a basis of $\cH_c$ and 
the dual base as $\psi^i$, 
where the degree of the dual basis is set by 
$\deg(\psi^i)=-\deg(\eb_{i})$. 
Then $\C[[\psi]]$ is the formal graded power series ring 
in the variables $\{\psi^i\}$. 

Let us express 
the $L_\infty$--structure $l_k \co  (\cH_c)^{\otimes k}\to\cH_c$, 
in terms of the bases: 
\begin{equation*}
 l_k(\eb_{i_1},\dots,\eb_{i_k})
 =\eb_{j}c^j_{i_1\cdots i_k}\ . 
\end{equation*}
Correspondingly, let us define an odd formal vector field 
on $\cH_c$, that is, a derivation of $\C[[\psi]]$:
\begin{equation}
 \delta_S =\flpartial{\psi^j}c^j(\psi)
 =\sum_{k\ge 0}\ov{k!}
 \flpartial{\psi^j}c^j_{i_1\cdots i_k}
 \psi^{i_k}\cdots\psi^{i_1}\ ,\qquad c^j(\psi)\in \C[[\psi]]\ . 
\end{equation}
Also, define a collection of multilinear structures on $\cH_o$ 
parameterized by $\cH_c$ as follows: 
\begin{equation}
 n_{i_1,\dots,i_p; q}(o_1,\dots, o_q)
 :=n_{p,q}(\eb_{i_1},\dots,\eb_{i_p}; o_1,\dots,o_q)\ 
\end{equation}
for $p,q\ge 0$ with $p+q>0$. 
Then, let us define a new collection of multilinear maps on 
$\ti{\cH_o}:=\cH_o\otimes \C[[\psi]]$ as follows: 
\begin{align}
  \ti{m}_n& :=\sum_{k\ge 0}
 n_{i_1\cdots i_p,q}\psi^{i_p}\cdots\psi^{i_1}\, 
 \co \, (\ti{\cH_o})^{\otimes n}\to \ti{\cH_o}
 \ ,\qquad (n\ne 1)\ ,\label{ti-m_n}\\
  \ti{m}_1 (\ti{o}) & :=\sum_{k\ge 0}
 n_{i_1\cdots i_p,q}\psi^{i_p}\cdots\psi^{i_1}(\ti{o})-\delta_S(\ti{o}), 
\label{ti-m_1}
\end{align}
where the tensor product $\ti{o}^{\otimes n}$ 
is defined over $\C[[\psi]]$. 
One can see that 
$(\ti{\cH_o},\{\ti{m}_k\}_{k\ge 0})$ 
forms a weak $A_\infty$--algebra over $\C[[\psi]]$. 
In fact, generalized to a more general base manifold ${\mathcal M}$, 
it may be plausible to call this a {\em geometric}
weak $A_\infty$--structure more general than Merkulov's,
which can be regarded as the case in which 
$\cH_o$ is the fiber of $T_0\cH_c$, the tangent space of $\cH_c$ 
at the origin of $\cH_c$ and then we drop the second condition 
of his geometric $A_\infty$--structure. 
Clearly an isomorphism of $T_0\cH_c$ to $\cH_c$ 
as graded vector spaces extends to an isomorphism 
from ${\mathcal T}_{\cH_c}$ to $\ti{\cH_c}:=\cH_c\otimes\C[[\psi]]$, 
cf \cite{mer1}, subsection 3.8.1. 
Under this identification, let us consider the 
particular case $n_{p,1}:=\ov{(p+1)!}l_{p+1}$. 
Thus, the differential $\ti{m}_1$ in equation \eqref{ti-m_1} turns out to be 
\begin{equation*}
 \ti{m}_1 (\ti{c})=-[\delta_S, \ti{c}],\qquad\ti{c}\in \ti{\cH_c}. 
\end{equation*}
Since he does not treat the `weak' case, $n_{p,0}$ is of course 
zero for any $p$. 
The higher multilinear maps 
$\ti{m}_n\co (\ti{\cH_c})^{\otimes n}\to\ti{\cH_c}$, 
$n\ge 2$, are defined in the same way as in equation \eqref{ti-m_n}, 
only with the replacement of elements in $\ti{\cH_o}$ 
by those in $\ti{\cH_c}$. 
One can see that $(\ti{\cH_c},\ti{m})$ obtained as above coincides with 
the geometric $A_\infty$--structure in subsection 3.8.1 of \cite{mer1}. 
Then, the theorem in subsection 3.8.2 in \cite{mer1} states 
that a geometric $A_\infty$--structure is equivalent to 
a ${\cal G}erst_\infty$--algebra structure which is defined by 
certain relations described by tree graphs having 
both straight edges and wiggly edges as in subsection 3.6.1 
of \cite{mer1}. 
Unfortunately, there a straight (resp. wiggly) edge 
corresponds to an element in $\ti{\cH_o}$ (resp. $\ti{\cH_c}$), 
so his convention is the opposite of ours for OCHAs. 
Even taking this into account, his defining equation 
for a ${\cal G}erst_\infty$--algebra structure 
is superficially different from ours. 
This is because we have $n_{p,1}=\ov{(p+1)!}l_{p+1}$ now; 
in the ${\cal G}erst_\infty$--algebra case, 
the $L_\infty$--structure $l_k$ (which is denoted by $\nu_k$ in \cite{mer1}) 
is our $l_k$ or our $k! n_{k-1,1}$. 
Remembering this fact, one can see that 
the ${\cal G}erst_\infty$--algebra condition 
in \cite{mer1} is a special case of our OCHA condition.

The commutative version of the geometric $A_\infty$--structure 
is called a geometric $C_\infty$--structure \cite{mer1,mer2}, 
which is a special $G_\infty$--algebra and plays an important role 
in deformation theory. 
For a given geometric $C_\infty$--structure, 
if we concentrate on the degree zero part 
of the graded vector space $\cH_c$, 
the higher products are also concentrated on the one with degree zero. 
The resulting special geometric $C_\infty$--structure is 
what is called an $F$--manifold, a generalization 
of a Frobenius manifold.

 \section{Applications of OCHAs to deformation theory}
\label{ssec:defs}

Consider
an OCHA $(\cH=\cH_c\oplus\cH_o,\l,\n)$. 
We will show how the combined structure implies 
the $L_\infty$--algebra $(\cH_c,\l)$ controls some
deformations of the $A_\infty$--algebra
$(\cH_o,\{m_k\}_{k\ge 1})$. We will further investigate the deformations
of this control as $\cH$ is deformed.

We first review some of the basics of deformation theory from a 
homotopy algebra point of view. 
The philosophy  of deformation theory which we follow 
(due originally, we believe, to Grothendieck
\footnote{See \cite{doran:bib} 
for an extensive annotated bibliography of deformation theory.}
 cf \cite{Sjds,goldman-millson,deligne}) 
regards any deformation theory as `controlled' 
by a dg Lie algebra $\g$ 
(unique up to homotopy type as an $L_\infty$--algebra).

For the deformation theory of an (ungraded) associative algebra
$(A,m)$ \cite{gerst:defm} the standard controlling dg Lie algebra
is $\Coder(T^c A)$ with the graded commutator as the graded Lie
bracket \cite{jds:intrinsic}.  Under the identification (including a
shift in grading) of $\Coder (T^c A)$ with $\Hom(T^cA, A)$ (which is
the Hochschild cochain complex), this bracket is identified with the
Gerstenhaber bracket and the differential with the Hochschild
differential, which can be written as $[m, \ ]$ \cite{gerst:coh}.

The generalization to a differential graded associative algebra is 
straightforward; the differential is now: $[d_A+m_2,\quad]$. 
For an $A_\infty$--algebra,
the differential similarly generalizes to $[\m,\ ]$. 

Deformations of $A$ correspond to certain elements of  
$\Coder (T^c A)$, namely 
those that are solutions of an {\em integrability} equation, 
now known more commonly as a {\em Maurer--Cartan} equation.
\begin{defn}[{The classical Maurer--Cartan equation}] 
In a dg Lie algebra $(\g,d,[\ ,\ ])$,
the {\em classical  Maurer--Cartan equation} is
\begin{equation}
 d\theta + \ov{2}[\theta,\theta]=0\qua \text{for}\ \theta\in \g^1.
 \label{mceq-dgla}
\end{equation}
\end{defn}
For an $A_\infty$--algebra $(A,\m)$ and $\theta \in \Coder^1(T^c A)$, 
a deformed $A_\infty$--structure 
is given by $\m + \theta$ iff 
\begin{equation*}
 (\m + \theta)^2 = 0\ .
\end{equation*}
Teasing this apart, since we start with $\m^2=0$, we have equivalently 
\begin{equation}
 D\theta + \ov{2}[\theta, \theta] = 0,
 \label{mceqCoder}
\end{equation}
the Maurer--Cartan equation of the dg Lie algebra 
$(\Coder(T^c A),D,[\ ,\ ])$ 
(Here $D$ is the natural differential on
$\Coder(T^c A)\subset \End (T^c A)$, ie $D\theta = [\m,\theta]$. )

For $L_\infty$--algebras, the analogous remarks hold, substituting
the Chevalley--Eilenberg complex for that of Hochschild, ie, using
$\Coder\ C(L) \simeq \Hom (C(L), L)$. 
\begin{defn}
[{The strong homotopy Maurer--Cartan equation}] 
In an $L_\infty$--algebra $(L,\l)$, 
the {\em (generalized) Maurer--Cartan equation} is
\begin{equation*}
 \sum_{k\geq 1} \ov{k!} l_k(\cb,\dots,\cb) = 0
\end{equation*}
for $\cb\in L^0$. 
\footnote{Note that the degree of $\cb$ is zero since 
a dg Lie algebra is precisely a special $L_\infty$--algebra 
after a suitable degree shifing called the {\em suspension} 
and then $\g^1=L^0$. }
\end{defn}
We denote the set of solutions of the Maurer--Cartan equation as 
$\cMC(L,\l)$ or more simply $\cMC(L)$.

Now, since an OCHA can be thought of as a generalization of 
an $A_\infty$--algebra over an $L_\infty$--algebra 
(\fullref{defn:AovL}), one has: 
\begin{thm}{\rm\cite{ocha,pocha}}\label{thm:ocha-Lmorp}\qua
An OCHA $(\cH:=\cH_c\oplus\cH_o,\l,\n)$ is equivalent to 
an $L_\infty$--morphism 
from $(\cH_c,\l)$ to $(\Coder(T^c(\cH_o)),D=[\m,\ ],[\ ,\ ])$, 
where $\m$ is the codiferential on $\Coder(T^c(\cH_o))$ corresponding 
to $\{m_k=n_{0,k}\}_{k\ge 1}$. 
\end{thm}
Since it is known that an $L_\infty$--morphism preserves 
the solutions of the Maurer--Cartan equations, we obtain the following: 
\begin{thm}
For an OCHA $(\cH:=\cH_c\oplus\cH_o,\l,\n)$, 
a Maurer--Cartan element $\cb\in\cM(\cH_c,\l)$ 
gives a deformation of $(\cH_o,\m:=\{n_{0,k}\}_{k\ge 1})$ 
as a {\em weak} $A_\infty$--algebra.
\end{thm}

For a dg Lie algebra, there is the notion of gauge transformation. 
A gauge transformation defines an equivalence relation $\sim$ between 
elements in $L$; two elements in $L$ are equivalent iff 
they are related by a gauge transformation. 
In particular, gauge transformations preserves 
the solution space $\cMC(L)$. 
Thus, the quotient space of $\cMC(L)$ by the equivalence relation 
$\sim$ is well-defined: 
\begin{equation*}
 \cM(L):=\cMC(L)/\sim. 
\end{equation*}
The moduli space of deformations 
is defined as this $\cM(L)$. 
In particular, one has isomorphic spaces 
$\cM(L)$ for any $L$ of the same $L_\infty$--homotopy type. 
Thus, deformation theory is in general 
controlled by an $L_\infty$ homotopy class of a dg Lie algebra. 

In general, to construct or even 
show the existence of an $L_\infty$--morphism 
is a very difficult problem. 
However, there exists a pair of a dg Lie algebra and an $L_\infty$--algebra 
where the existence of an $L_\infty$--morphism between them is 
guaranteed in some sense. 
We shall explain this below.

 \section{Homotopy algebra and string theory}
\label{sec:6}
Let us start from a `definition' of string theory 
which provides the main motivation
 for obtaining a pair of a dg Lie algebra and an 
$L_\infty$--algebra together with an $L_\infty$--morphism between them. 

A string is a one dimensional object whose `trajectory' (worldsheet) 
is described as a 
Riemann surface (two-dimensional object). 
In string theory, the most fundamental quantities are the scattering 
amplitudes of strings. 
The most general Riemann surfaces to be concerned with are 
those with genera, boundaries, and punctures on the boundaries and/or 
in the interior, where a puncture on a boundary (resp. in the interior) 
corresponds to an open (resp. closed) string insertion. 
For a fixed appropriate field theory on  Riemann surfaces, 
the scattering amplitudes are obtained by choosing a suitable 
compactification of the moduli spaces of punctured Riemann surfaces; 
the scattering amplitudes are integrals over the 
compactified moduli spaces. 
The collection of the scattering amplitudes obtained as above 
are endowed with special algebraic structures, 
associated to the stratifications of the compactifications 
of the moduli spaces. 
In this sense,
a definition of a string theory is the pair of an operad 
(associated to the compactified moduli spaces) 
and a representation of the operad (an algebra over the operad). 

In particular, if we consider a {\em sigma-model} on a Riemann surface, 
\ie, 
a field theory whose fields are maps from the Riemann surfaces 
to a target space $M$, the representation obtained by the field
theory has some information about the geometry of $M$.

As we explained, 
an $A_\infty$--algebra $(A,\m)$ is obtained as a representation 
of the $A_\infty$--operad $\cA_\infty$ on $A$. 
So a deformation of $(A,\m)$ 
is a deformation of a representation 
of the $A_\infty$--operad $\cA_\infty$ on the fixed graded vector space $A$. 

Here, recall that the $A_\infty$--operad $\cA_\infty$ is a structure 
which is associated to the real compactification of the moduli spaces of 
disks with punctures on the boundaries. 

On the other hand, 
a representation of the $L_\infty$--operad $\cL_\infty$ 
on a fixed graded vector space $L$ is 
an $L_\infty$--algebra $(L,\l)$. 

Now, if we fix a tree open string theory 
and a tree closed string theory on Riemann surfaces, 
we obtain an 
$A_\infty$--algebra $(\cH_o,\m)$ and the 
dg Lie algebra $(\Coder(T^c(\cH_o)),$\break $D, [\ ,\ ])$ controlling its deformation 
for the tree open string theory, 
and also an $L_\infty$--algebra $(\cH_c,\l)$ for the tree closed string. 
Moreover, the tree open--closed string system provides 
a representation of  the $\cOC_\infty$--operad on $\cH:=\cH_c\oplus\cH_o$,
that is, an OCHA $(\cH,\l,\n)$. 
Together with \fullref{thm:ocha-Lmorp}, 
by considering an appropriate tree open--closed string (field) theory, 
one can get a pair of a dg Lie algebra 
and an $L_\infty$--algebra 
together with a non-trivial $L_\infty$--morphism betweeen them.

One example of this is  Kontsevich's set-up \cite{kont:defquant-pub} for 
deformation quantization \cite{bfflsI,bfflsII} (see \cite{pocha}). 
Furthermore, as $L_\infty$--algebras $\cH_c$ of 
closed string (field) theories, 
one can consider for instance the dg Lie algebra 
controlling the extended deformation of complex structures \cite{barakont} 
(which is what is called the B--model in string theory (cf \cite{pocha})), 
or the dg Lie algebra controlling deformation 
of generalized complex structure \cite{gual}, etc.

Then, OCHAs should guarantee the existence 
of corresponding deformations of the $A_\infty$--structures. 
Currently, it should be a very interesting problem 
to descibe explicitly such a deformation of an $A_\infty$--structure 
as was done in the case of $*$--product for deformation 
quantization \cite{kont:defquant-pub}. 
Some attempts in this direction can be found in 
\cite{Hof2} for B--twisted topological strings and 
\cite{pestun} for the case of generalized complex structures. 

In these situations, which are related to mirror symmetry, 
the $L_\infty$--algebra corresponding to the tree closed string theory 
is (homotopy equivalent to) trivial, which implies 
that the deformation of the corresponding $A_\infty$--structure 
is unobstructed (see \cite{pocha}). 
In such situations, deformation of $A_\infty$--structures can be 
thought of as a (homological) algebraic description 
of an $L_\infty$--algebra $(\cH_c,\l)$ describing deformation of a geometry 
(see homological mirror symmetry by Kontsevich \cite{HMS}). 

\subsubsection*{Acknowledgments}
In working on OCHA, we had had valuable discussions 
with 
E~Harrelson, A~Kato, T~Kimura, T~Lada, M~Markl,   
S\,A~Merkulov, H~Ohta, A~Voronov and B~Zwiebach.

JS is supported in part by NSF grant FRG DMS-0139799 
and US--Czech Republic grant INT-0203119.

\bibliographystyle{gtart}
\bibliography{link}

\end{document}